\documentclass[aps,prl,twocolumn,notitlepage,nofootinbib,longbibliography]{revtex4-2}

\usepackage[dvips]{graphicx}
\usepackage{braket}
\usepackage{dcolumn}
\usepackage{comment}
\usepackage{bm}
\usepackage{amsmath,amsthm,amssymb}
\usepackage{epstopdf}
\usepackage{mathrsfs}
\usepackage{color}
\usepackage{xcolor}
\usepackage{bbold}
\usepackage{mathtools}
\usepackage[sort&compress]{natbib}
\usepackage[pdftex,colorlinks=true,citecolor=black,urlcolor=black,linkcolor=black]{hyperref}

\definecolor{azure}{rgb}{0.0, 0.5, 1.0}
\definecolor{amber}{rgb}{1.0, 0.49, 0.0}
\definecolor{red}{rgb}{1.0, 0.1, 0.0}
\definecolor{forestgr}{rgb}{0.13, 0.55, 0.13}

\newcommand{\+}{^\dagger}
\newcommand{\ha}{\hat a}
\newcommand{\hb}{\hat b}
\newcommand{\hc}{\hat c}
\newcommand{\Rev}[1]{{\color{black}#1}}

\begin{document}

\title{Squeezing Multilevel Atoms in Dark States via Cavity Superradiance}

\author{Bhuvanesh Sundar}
\thanks{Present address: Rigetti Computing, Berkeley, California 94710, USA}
\author{Diego Barberena}
\author{Ana Maria Rey}
\author{Asier Pi\~neiro Orioli}

\affiliation{JILA, NIST, Department of Physics, University of Colorado, Boulder, Colorado 80309, USA}
\affiliation{Center for Theory of Quantum Matter, University of Colorado, Boulder, Colorado 80309, USA}

\date{\today}

\begin{abstract}
We describe a method to create and store scalable and long-lived entangled spin-squeezed states within a manifold of many-body cavity dark states using collective emission of light from multilevel atoms inside an optical cavity. 
We show that the system can be tuned to generate squeezing in a dark state where it will be immune to superradiance. We also show more generically that squeezing can be generated using a combination of superradiance and coherent driving in a bright state, and subsequently be transferred via single-particle rotations to a dark state where squeezing can be stored. Our findings, readily testable in current optical cavity experiments with alkaline-earth-like atoms, can open a path for dissipative generation and storage of metrologically useful states in optical transitions.
\end{abstract}

\maketitle

Subradiant states that emit light at a rate slower than independent atoms because of (quantum) interference~\cite{dicke1954coherence}, have attracted widespread interest owing to their potential applications in quantum memories~\cite{asenjo2017exponential}, mirrors~\cite{shahmoon2017cooperative, Rui2020}, excitation transport~\cite{needham2019subradiance, ballantine2020subradiance}, topological physics~\cite{perczel2017topological, perczel2017photonic, bettles2017topological, zhang2019tunable, syzranov2016emergent}, entangled photons~\cite{gonzalez2015deterministic}, and quantum metrology~\cite{henriet2019critical, Chu2019,Pineiro2019,Pineiro2020}.
A long-standing challenge is finding simple ways to prepare target many-body subradiant states with useful properties such as scalable entanglement, i.e. entanglement which increases with system size. 
Optical cavities have demonstrated the capability to create \emph{collective} (i.e., fully symmetric) quantum many-body states with scalable entanglement in the form of squeezing~\cite{ma2011quantum, schleier2010squeezing, leroux2010implementation, chen2014cavity,Cox2016,hosten2016measurement,Pezze2018}. \Rev{However, creating optically excited entangled states that are immune to collective dissipation and metrologically useful has remained a major challenge.}

In generic atom-cavity experiments with two-level atoms, collective states are typically not dark but superradiant~\cite{dicke1954coherence,GrossHarocheSuperr,norcia2016superradiance}.
One way to stabilize the decay and create scalable entanglement is by using an additional coherent drive which competes with superradiance~\cite{barberena2019driven, wolfe2014spin, dalla2013dissipative, walls1978non, drummond1978volterra, drummond1980multiple, drummond1980observables, carmichael1980analytical, puri1979exact, somech2022quantum}.
However, after turning off the drive, excited atoms superradiantly decay to the ground state, and the entanglement is destroyed.

\begin{figure}[t]\centering
\includegraphics[width=1\columnwidth]{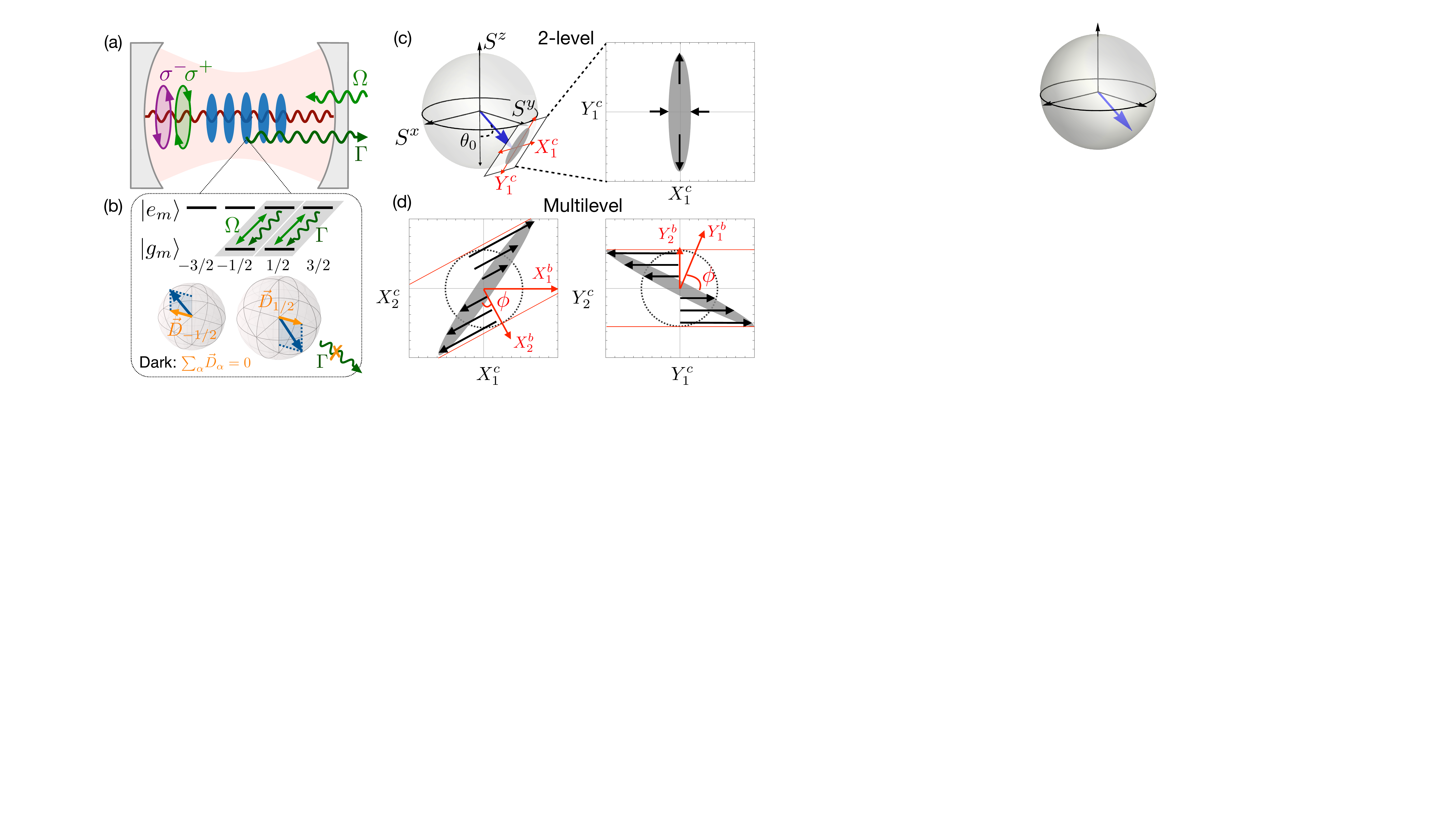}
\caption{(a) Atom-cavity setup: Atoms loaded inside a standing-wave cavity, resonant with the atomic transition, decay at a rate $\Gamma$. Atoms are coherently driven with right-circularly-polarized light with an effective Rabi frequency $\Omega$. (b) Destructive interference of the collective dipoles $\vec{D}_{\pm 1/2}$ among two transitions in six-level atoms leads to a dark state which does not decay to the ground state. Blue and orange arrows are the mean Bloch vector and dipole for the two transitions. (c) Squeezing in a two-level system, visualized on a Bloch sphere (left) and bosonic noise distribution (right). The steady state is squeezed along $\hat X^c_1 \propto \hat S^x$. (d) Squeezing in a multilevel system. The panels plot projections of the noise distribution onto Schwinger boson quadratures $\hat X^c_{1,2}$ and $\hat Y^c_{1,2}$, which are proportional to spin variables perpendicular to the multilevel Bloch vector. Black arrows indicate that the distributions shear perpendicular to conserved Bogoliubov bosonic variables $\hat X^b_2$ and $\hat Y^b_2$ (red arrows), which leads to two squeezed modes.
}
\label{fig1}
\end{figure}

Here, we propose to use \emph{multilevel} atoms coupled to a dissipative cavity [see Fig.~\ref{fig1}(a)] to generate scalable squeezing in \Rev{two distinct modes} and store it in dark states, recently shown to exist in these systems~\cite{orioli2022emergent, lin2022dissipation, fan2023collective}. 
At the mean-field level, these dark states can be understood as the cancellation of the collective dipoles corresponding to different internal atomic transitions, see orange arrows in Fig.~\ref{fig1}(b).
In the full quantum theory, however, these dark states are necessarily entangled, as revealed by their squeezed fluctuations, and are hence absent in single atoms.
We focus on effective four-level atoms [Fig.~\ref{fig1}(b)] and describe two cases: one where collective decay generates squeezing in a dark state directly, and one where squeezing is created in a bright state due to a combination of superradiance and coherent driving.
For the bright state, we show how to transfer the squeezing into a dark state using single-particle rotations, and store it by using symmetries which preserve quantum noise.
The storage protocol can in principle be applied to squeezing generated in multilevel systems via other schemes~\cite{norris2012enhanced, kurucz2010multilevel, masson2017cavity, Pezze2018,masson2017cavity,sundar2023bosonic} such as near-unitary two-mode squeezing in cavities.

The multilevel squeezing can be understood by approximating the spin projections orthogonal to the mean collective spin, to bosonic degrees of freedom (large $N$ expansion). In this picture, multilevel superradiance leads to two-mode squeezing [Fig.~\ref{fig1}(d)], in contrast to the one-mode squeezing of two-level atoms [Fig.~\ref{fig1}(c)]. \Rev{This provides an alternative method to produce squeezed states in two independent modes, akin to the two-mode squeezed states realized in BECs and thermal gases~\cite{gross2011atomic,lucke2011twin,bookjans2011,Black2007,zhao2014,qu_2020,Kim2021, julsgaard2001experimental, polzik2016entanglement}.} 
Some spin projections are unaffected by superradiance, as suggested by the black arrows in Fig.~\ref{fig1}(d), where two specific bosonic quadratures [$X^b_2$ and $Y_2^b$ (red)] are conserved, a property absent from the two-level case. The squeezing can therefore be 
preserved by rotating the state such that the squeezed quadratures are aligned along the conserved directions.

\emph{Setup.---}
We consider an ensemble of $N$ atoms pinned in a deep magic optical lattice within an optical cavity [see Fig.~\ref{fig1}(a)]. For concreteness, we consider atoms with degenerate ground states $\ket{g,m_g}$, $m_g \in [-\frac{1}{2}, \frac{1}{2}]$, and long-lived electronic excited states $\ket{e,m_e}$, $m_e \in [-\frac{3}{2}, \frac{3}{2}]$ and ground-excited transition frequency $\omega_a$, where the quantization axis is along the cavity axis. 
This can be realized, for example, with the $^1S_0$ and $^3P_1$ states of $^{171}$Yb~\cite{ludlow2015optical}. While our arguments work for generic multilevel atoms, this is the simplest nontrivial atomic structure which shows the relevant physics. 

Two cavity modes with angular frequency $\omega_c$ and orthogonal polarizations couple ground state atoms to the excited state. For simplicity, we consider that the cavity is resonant with the atoms, $\omega_c = \omega_a = \omega$. 
We drive the cavity modes with a right-circularly-polarized laser whose frequency is also resonant with the cavity, $\omega_L = \omega$. In the bad cavity limit where the cavity field can be adiabatically eliminated~\cite{ SM}, the atom dynamics are described by the Lindblad master equation $\hbar\dot{\rho} = -i[\hat{H},\rho] + \mathcal{L}[\rho]$ with
\begin{align}
 \hat H =&\, \hbar\Omega \hat D^x_{+1}, 
 \label{eq:rabi_H}\\
 \mathcal{L}[\rho] =&\, \hbar\Gamma \sum_{\alpha=\pm 1} \Big( \hat{D}^-_\alpha \rho \hat{D}^+_\alpha - \frac{1}{2} \{\hat{D}^+_\alpha \hat{D}^-_\alpha, \rho \} \Big),
\label{eq:lindblad_general}
\end{align}
where $\{\cdots\}$ is the anticommutator. Equation~\eqref{eq:rabi_H} describes a right-circularly-polarized ($\alpha\,{=}\,{+}\,1$) coherent drive, 
and Eq.~\eqref{eq:lindblad_general} describes superradiant emission of left-($\alpha\,{=}\,{-}\,1$) and right-circular ($\alpha\,{=}\,{+}\,1$) polarizations. 
Here, $\Omega$ is the atoms' effective Rabi coupling strength generated by the drive,
and $\Gamma$ sets the rate at which atoms decay and emit light which leaks out of the cavity. 
The operator $\hat{D}^+_\alpha = \sum_{m} C_m^\alpha \hat S^+_{m,\alpha} $ describes the collective atomic excitation due to absorbing an $\alpha$-polarized photon that imparts $\alpha$ units of angular momentum, $\hat{S}^+_{m,\alpha} = \sum_{i=1}^N \ket{e,m+\alpha}_i \bra{g,m}_i$ is the collective spin-raising operator for $N$ atoms within the two-level manifold of $\ket{g,m}_i$ and $\ket{e,m+\alpha}_i$, $C_m^\alpha = \braket{ F_g, m ; 1, \alpha \vert F_e, m+\alpha }$ is the Clebsch-Gordan coefficient for the associated transition, 
and $m$ is the angular momentum projection of $\ket{g,m}$ onto the quantization axis.
We further define $\hat D^x_\alpha = [(\hat D^+_\alpha + \hat D^-_\alpha)/2]$ and $\hat D^y_\alpha = [(\hat D^+_\alpha - \hat D^-_\alpha)/2i]$.

We initialize the atoms in a product of single-particle ground states $\ket{G_\beta} = \cos(\beta/2) \ket{g,-1/2} + \sin(\beta/2) \ket{g,1/2}$, and apply a right-circularly-polarized pulse to excite a fraction of the atoms to the excited state, leaving them in the coherent state
$\ket{\Psi_{\theta_0;\beta}} = \ket{\psi_{\theta_0;\beta}}^{\otimes N} = \exp( -i\theta_0 \hat D^x_{+1} ) \ket{G_\beta}^{\otimes N}$. 
The levels $\ket{e,-3/2}$ and $\ket{e,-1/2}$ are always empty, so only the right-handed polarization is relevant [Fig.~\ref{fig1}(b)]. For convenience, we define a family of states $\ket{\Psi(\theta;\beta)} = \exp( -i\theta \hat D^x_{+1}) \ket{G_\beta}^{\otimes N}$, with $\ket{\Psi_{\theta_0;\beta}} = \ket{\Psi(\theta_0;\beta)}$.

To study the subsequent evolution it is useful to rewrite the above master equation as $\hbar\dot{\rho} = \mathcal{L}'[\rho]$ where $\mathcal{L}'[\rho]$ looks like $\mathcal{L}[\rho]$ [Eq.~(\ref{eq:lindblad_general})] but with the modified jump operators
\begin{equation}
\hat{\mathscr{D}}^-_\alpha = \hat D^-_\alpha + i \frac{\Omega_\alpha}{\Gamma},
\end{equation}
where $\Omega_{+1} = \Omega$ and $\Omega_{-1} = 0$.
One can then show that the system's steady state fulfills $\hat{\mathscr{D}}^-_{+1}\rho_{\rm ss} = 0$~\cite{puri1979exact, somech2022quantum, barberena2019driven}.

\textit{Mean-field physics.---}
A convenient way to visualize the state under the mean-field approximation is in terms of multiple Bloch spheres (labeled $m$), one for each of the two-level subspaces composed of $\ket{g,m}$ (south pole) and $\ket{e,m+1}$ (north pole), see Fig.~\ref{fig1}(b).
The dynamics of the $m$th Bloch vector, $\vec{S}_{m} = (\braket{ \hat S^x_{m}}, \braket{\hat S^y_{m}}, \braket{\hat S^z_{m}})$, where expectation values are taken in the mean-field state and we dropped the subscript $\alpha$, is given by~\cite{SM} $\dot{\vec{S}}_m = C_m^{+1}(\vec{\Omega} + \Gamma\vec{D}_\perp) \times \vec{S}_m$ where $\vec{\Omega} = (\Omega, 0, 0)$ and 
$\vec{D}_\perp = (-\braket{\hat{D}^y_{+1}}, 0, 0)$ 
 for the initial state $\ket{\Psi_{\theta_0;\beta}}$.
Therefore, both the Rabi drive and superradiance separately lead to a rotation of each Bloch vector around an axis with fixed directions $\vec{\Omega}/|\vec{\Omega}|$ at a rate $\propto C^{+1}_m$. 
This means that all Bloch vectors can be described by the single angle $\theta(t)$ as $\vec{S}_m(t) = |\vec{S}_m| (0, \sin[C_m^{+1} \theta(t)], -\cos[C_m^{+1} \theta(t)])$, where $|\vec{S}_m|$ is constant and $\theta(0)=\theta_0$. The mean-field time-evolved state associated with this Bloch vector is $\ket{\Psi\big(\theta(t);\beta\big)}$.

\begin{figure}[t]\centering
\includegraphics[width=1\columnwidth]{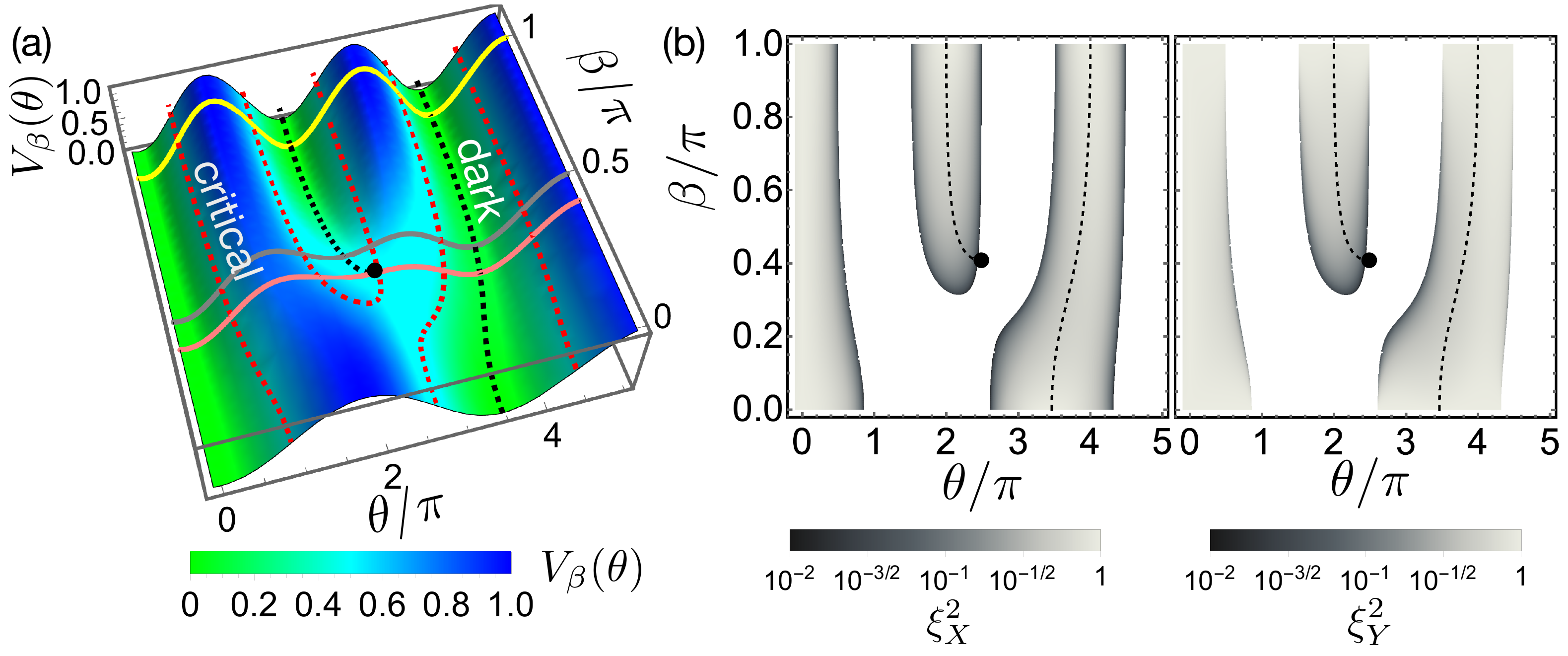}
\caption{Superradiance potential $V_\beta(\theta)$. 
Both height and color indicate $V_\beta(\theta)$, black dashed lines are stable mean-field dark states, and red dashed lines are critical lines. Pink, gray, and yellow lines are parametric cuts explored in this Letter. (b) Steady-state squeezing $\xi^2_{X(Y)}$ in the $\hat X$ ($\hat Y$) quadratures. The squeezing becomes significant near the critical points [red dashed lines in (a)]. White regions are unstable. 
The dot indicates where the dark state and critical point coincide.}
\label{fig2}
\end{figure}

The angle $\theta(t)$ evolves according to $\dot\theta = \Omega - N\Gamma [\partial V_\beta(\theta)/\partial\theta]$, where $V_\beta(\theta) = \frac{1}{2} + \frac{1}{N} \braket{ \Psi(\theta;\beta) \vert \sum_m \hat{S}^z_m \vert \Psi(\theta;\beta)}$ is the superradiance potential~\cite{orioli2022emergent}. 
The potential $V_\beta(\theta) = \cos^2(\beta/2)\sin^2(\theta/2\sqrt{3}) + \sin^2(\beta/2)\sin^2(\theta/2)$ is plotted in Fig.~\ref{fig2}(a). 
We can visualize the mean-field dynamics as the classical evolution of a particle with coordinate $\theta$ on the potential $V_\beta(\theta)$. Superradiance, which is a form of dissipation, pulls $\theta$ towards a minimum of the potential, keeping $\beta$ fixed, whereas the coherent drive increases (decreases) $\theta$ at a constant rate for $\Omega >0\ (\Omega<0)$.

The mean-field steady state ($\dot\theta=0$) is given by the solution to $\Omega = N\Gamma [\partial V_\beta(\theta)/\partial\theta]$, when it exists. The stability of this steady state is determined by the curvature of the potential. 
The steady state is stable to quantum fluctuations for $[\partial^2V_\beta(\theta)/\partial\theta^2]>0$, and unstable for $[\partial^2V_\beta(\theta)/\partial\theta^2]<0$. The inflection points at $[\partial^2V_\beta(\theta)/\partial\theta^2]\vert_{\theta=\theta_c} = 0$ [red dashed lines in Fig.~\ref{fig2}(a)] determine critical lines that separate the stable and unstable regions. In the absence of drive, $\Omega = 0$, the steady states are mean-field dark states, and occur when the potential has a minimum with respect to $\theta$ [black dashed lines in Fig.~\ref{fig2}(a)]. They emerge from destructive interference of the collective dipoles $\vec{D}_m$ in the various Bloch spheres [Fig.~\ref{fig1}(b)]~\cite{orioli2022emergent}.

The steady-state solution for $\theta$ depends on $\theta_0$, $\Omega/N\Gamma$, and $\beta$. For simplicity, we choose throughout the Letter
$\Omega = N\Gamma [\partial V_\beta(\theta)/\partial\theta]\vert_{\theta=\theta_0}$ and $[\partial^2V_\beta(\theta)/\partial\theta^2]\vert_{\theta=\theta_0} > 0$. 
For this choice, the initial state $\ket{\Psi_{\theta_0;\beta}}$ is a stable mean-field steady state. Below the critical points, $\ket{\Psi_{\theta_0;\beta}}$ is a good approximation to the full quantum steady state, and we can treat fluctuations around the mean field as a perturbation.

\textit{Quantum noise.---}
We treat the perturbations around the mean field in a bosonic picture in the large-$N$ approximation. For atoms with $\ell$ relevant internal levels in the ground and excited manifolds, we can define Schwinger boson operators $\ha_{g(e),m}$ that annihilate particles in $\ket{g(e),m}$, but a more convenient choice turns out to be $\theta_0$-dependent operators $\hc_\mu(\theta_0,\beta)$, $\mu\in [0, \ell-1]$, that are related to $\ha_{g(e),m}$ via a unitary transformation. 
For brevity, we will drop the $(\theta_0,\beta)$ dependence of $\hc_\mu$.

We choose the definition of $\hc_\mu$ based on two considerations. First, we define $\hc_{\mu=0}$ to annihilate particles in $\ket{\psi_{\theta_0;\beta}}$. 
Since we have chosen the initial state $\ket{\Psi_{\theta_0;\beta}}$ to be a mean-field steady state, 
in the large-$N$ limit we can make the generalized Holstein-Primakoff (HP) approximation~\cite{kurucz2010multilevel}, $\hc_0 \sim \sqrt{N}$, and assume that the occupation in $\ket{\psi_{\theta_0;\beta}}$ is always close to $N$.
Note that $\Ket{\Psi_{\theta_0;\beta}}$ is a macroscopically occupied state of $\hc_0$, and the coherent vacuum for $\hc_{\mu>0}$. The real and imaginary parts of $\hc_{\mu>0}$, $\hat X^c_\mu = [(\hc_\mu^{\phantom\dagger} + \hc_\mu\+)/\sqrt{2}]$ and $\hat Y^c_\mu = [(\hc_\mu^{\phantom\dagger} - \hc_\mu\+)/\sqrt{2}i]$, describe spin variables perpendicular to the Bloch vector (see below). 
Second, we define $\hc_{\mu>0}$ such that $\hat{\mathscr{D}}^-_{+1}$, which is a linear combination of $\hc_{\mu>0}$, has a simple form~\cite{SM}. We exemplify this later.

We diagonalize the Lindbladian with a Bogoliubov transformation of $\hc_{\mu>0}$ which defines Bogoliubov bosons $\hb_{\mu>0}$ such that $\hat{\mathscr{D}}^-_{+1} \propto \sqrt{N} \, \hb_1$. 
$\hat{b}_\mu$ also depends on $(\theta_0,\beta)$ and we denote their real and imaginary parts as $\hat X^b_\mu$ and $\hat Y^b_\mu$.
This transformation lets us visualize the many-body steady state satisfying $\hat{\mathscr{D}}^-_{+1}\rho_{\rm ss} = 0$, as the vacuum of $\hb_1$. 
Thus, during the dynamics, the system evolves from the vacuum of $\hc_{\mu>0}$, which has a positive value for $\braket{ \hb_1\+ \hb_1} = \frac{1}{2} [(\Delta X^b_1)^2 + (\Delta Y^b_1)^2 - 1]$, to the vacuum of $\hb_1$, which has $\braket{ \hb_1\+ \hb_1} = 0$ and $(\Delta X^b_1)^2 = (\Delta Y^b_1)^2 = 1/2$, while the $\hat{b}_{1<\mu< \ell}$ bosons remain untouched [Fig.~\ref{fig1}(d)]. 
This noise reduction perpendicular to the Bloch vector corresponds to the generation of spin squeezing.

Specifically, the system is squeezed if some of the spin variables $S_{\perp,\gamma}$ perpendicular to the Bloch vector fulfill $\xi_\gamma^2 \equiv 4(\Delta S_{\perp,\gamma})^2/N < 1$. In our approximation, this corresponds to a variance in the $\hc_{\mu>0}$ bosons. We find the smallest variances by calculating the $2(\ell-1)$ eigenvalues $\xi_\gamma^2$ of the covariance matrix $\Sigma$, whose matrix elements are 
$\{ \braket{\{\hat X^c_\mu, \hat X^c_\nu\}}_{C}, \braket{\{\hat X^c_\mu, \hat Y^c_\nu\}}_{C}, \braket{\{\hat Y^c_\mu, \hat Y^c_\nu\}}_{C}\}$, where $\braket{\{\hat A,\hat B\}}_{C} = \braket{\hat A\hat B + \hat B\hat A}-2\braket{\hat A}\braket{\hat B}$~\cite{SM}.
Initially, $\Sigma$ is the identity matrix, which corresponds to no squeezing. As the driven-dissipative system evolves, some eigenvalues of $\Sigma$ become squeezed, $\xi_\gamma^2 <1$. We show in~\cite{SM} that an eigenvalue $\xi_\gamma^2 <1$ is an entanglement witness~\cite{vitagliano2011spin, vitagliano2014spin}.

\textit{Two-level systems.---}
To exemplify squeezing in the simplest case, we consider $\ket{G_{\beta=\pi}} = \ket{g,1/2}$ [yellow line in Fig.~\ref{fig2}(a)]. In this case, the dynamics is constrained to $\ket{g,\frac{1}{2}}$ and $\ket{e,\frac{3}{2}}$, effectively realizing a two-level system whose physics has been extensively studied previously~\cite{walls1978non, drummond1978volterra, drummond1980multiple, drummond1980observables, carmichael1980analytical, puri1979exact, barberena2019driven, wolfe2014spin, somech2022quantum}. 
For this case, the superradiance potential is $V_{\beta=\pi}(\theta) = \sin^2(\theta/2)$, whose critical points are $\theta_c = \pm (\pi/2)$. The steady-state Bloch vector is along $\hat{S}^{\rm Bloch}_{\theta_0} = \sin\theta_0 \hat S^y_{1/2} - \cos\theta_0 \hat S^z_{1/2}$, stabilized at the mean-field level with a drive strength $\Omega = (N\Gamma/2)\sin\theta_0$ for $|\theta_0|<(\pi/2)$. 
Defining Schwinger bosons such that~\cite{SM} $\hat X^c_1 \approx \sqrt{(2/N)} \hat S^x_{1/2}$ and $\hat Y^c_1 \approx \sqrt{(2/N)}(\cos\theta_0 \hat S^y_{1/2} + \sin\theta_0 \hat S^z_{1/2})$ in the HP approximation [see Fig.~\ref{fig1}(c)],
the Lindblad operator can be written as $\hat{\mathscr{D}}^-_{+1} \approx \sqrt{(N/2)}(\hat X^c_1 + i\cos\theta_0\hat Y^c_1)$.
Thus, we define the Bogoliubov operator $\hb_1 = \hat (X^c_1/\sqrt{2\cos\theta_0}) + i \hat Y^c_1\sqrt{(\cos\theta_0/2)}$, so that $\hat{\mathscr{D}}^-_{+1} = \sqrt{N\cos\theta_0}\ \hb_1$. 
Since the steady state is the vacuum of $\hb_1$ with 
$(\Delta \hat{X}^b_1)^2 = (\Delta \hat{Y}^b_1)^2 = \frac{1}{2}$, this implies that $\hat X^c_1$ and $\hat Y^c_1$ are squeezed and antisqueezed in the steady state, respectively, with $(\Delta X^c_1)^2 = (\cos\theta_0/2)$ and $(\Delta Y^c_1)^2 = (1/2\cos\theta_0)$ as shown in Fig.~\ref{fig1}(c).
The squeezing (antisqueezing) approaches 0 ($\infty$) as $\theta_0$ approaches the critical point, $\theta_c = \pm(\pi/2)$.

\textit{Multilevel systems.---}
When both ground levels are initially populated, the system hosts four nontrivial Schwinger bosons $\hc_\mu$ and thus three Bogoliubov bosons $\hb_\mu$ . We define $\hc_\mu$ such that the jump operator is~\cite{SM}
\begin{align}\label{eqn: Dminus HP}
\hat{\mathscr{D}}^-_{+1} & \approx \sqrt{N}\left(x \hat X^c_1 + i y (\cos\phi \hat Y^c_1 + \sin\phi \hat Y^c_2)\right),
\end{align}
where $(x, y, \phi)$ are parameters that depend on $(\theta_0,\beta)$. Specifically, \emph{all} the critical points, $[\partial^2V_\beta(\theta)/\partial\theta^2]\vert_{\theta=\theta_0} = 0$, correspond to $\phi = (\pi/2)$. As before, we define $\hb_1$ via $\hat{\mathscr{D}}^-_{+1} \approx \sqrt{N x y \cos\phi}\ \hb_1$. 
We define the other two Bogoliubov operators as $\hb_2 = \{[(\hat X^c_2 - \tan\phi \hat X^c_1) + i \hat Y^c_2]/\sqrt{2}\}$ and $\hb_3 = \hc_3$. These modes commute with $\hat{\mathscr{D}}^\pm_{+1} (\propto \hat b_1^{\phantom\dagger}, \hat b_1\+)$, and therefore their quadratures are conserved during dynamics. $\hb_2$ and $\hb_3$ are said to be generators of strong symmetries~\cite{lieu2020symmetry, prosen2008third, buvca2012note, barthel2022solving}. 

The evolution relaxes the system to the vacuum of $\hb_1$, which leads to dynamics in $\hc_1$ and $\hc_2$ only. The $\hb_3$ boson thus plays no role in the dynamics. However, since $\hb_2$ is conserved, the dynamics in $\hc_1$ and $\hc_2$ is such that the noise distributions shear 
perpendicular to $\hat X^b_2$ and $\hat Y^b_2$, as shown in Fig.~\ref{fig1}(d). 
This shearing leads to squeezing in \emph{two distinct modes} in the $\hat{c}$ basis, one in the $X^c_1$-$X^c_2$ plane and one in the $Y^c_1$-$Y^c_2$ plane, as opposed to the single squeezed mode of the two-level system above.
The shearing is reminiscent of spin squeezing via e.g. one-axis twisting (OAT). 
Unlike OAT, however, the dissipative dynamics does not preserve the area of the noise distribution, and also the shearing rate is time dependent and stops when $(\Delta X^b_1)^2 = (\Delta Y^b_1)^2 = \frac{1}{2}$.

Figure~\ref{fig2}(b) plots the squeezing in the two squeezed modes. 
The squeezing in both modes approaches 0 at the critical lines ($\phi=\pi/2$), but finite-$N$ effects limit the best squeezing achievable. Near the critical point, the squeezed quadratures approach $\hat X^c_1$, which is $\propto \hat D^x_{+1}$, and $\hat Y^c_2$, which is $\propto \hat D^y_{+1} - \braket{\hat D^y_{+1}}$~\cite{SM}. 

\begin{figure}[t]\centering
\includegraphics[width=1\columnwidth]{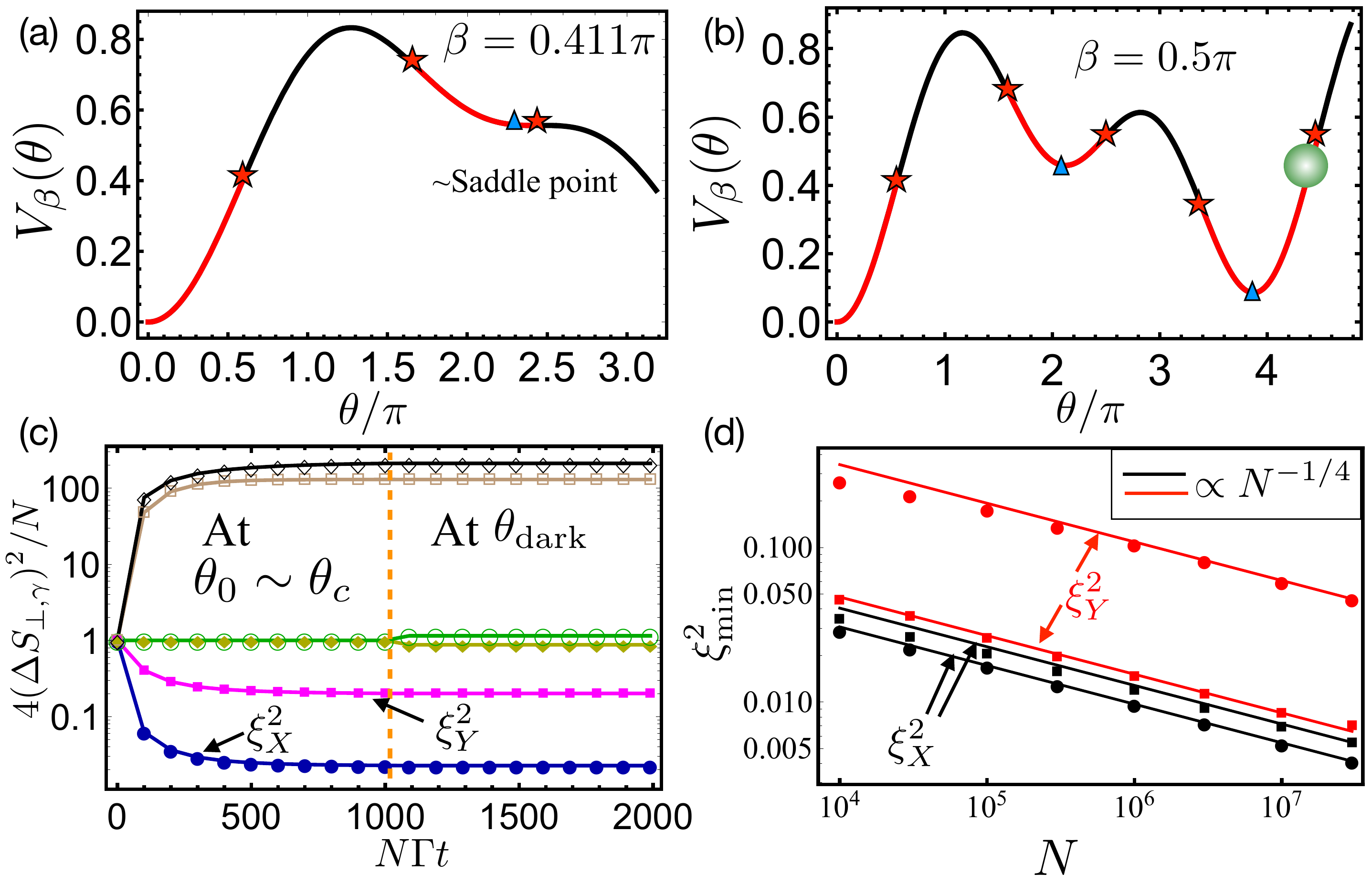}
\caption{
(a),(b) Superradiance potential $V_\beta(\theta)$ at (a) $\beta = 0.41\pi$ and (b) $\beta = 0.5\pi$. Blue triangles and red stars indicate dark states and critical points respectively.
(c) The six leading eigenvalues of the spins' covariance matrix $\tilde\Sigma$ during the storage protocol at $(\theta_0, \beta) = (4.46\pi, 0.5\pi)$, calculated using a cumulant expansion truncated at second order~\cite{SM}.
These eigenvalues correspond to the six normalized variances, $4(\Delta S_{\perp,\gamma}^2)/N$, of the spin variables perpendicular to the Bloch vector. 
The two squeezed modes preserve the noise after rotating to the dark state at $t = 1000/(N\Gamma)$. Lines are guides to the eye. (d) Finite-size scaling of the best squeezing in the $\hat X$ and $\hat Y$ quadratures near the critical points $(\theta_c,\beta) = (4.47\pi,0.5\pi)$ [dots] and $(\theta_{c,\rm dark},\beta_{c,\rm dark}) = (2.45\pi,0.41\pi)$ [squares].}
\label{fig3}
\end{figure}

\emph{Squeezing in a dark state.---}
The simplest way to generate scalable squeezing in a dark state is to initially prepare the system in a mean-field dark state that is close to a critical point, and let the system evolve with $\Omega=0$.
We achieve this by choosing $\beta$ and $\theta_0$ appropriately.
Figure~\ref{fig2}(a) shows that the dark manifold (black dashed line) intersects the critical manifold (red dashed line) at a saddle point (black dot) given by $(\theta_{c,\rm dark}, \beta_{c,\rm dark}) \approx (2.45\pi, 0.41\pi)$~\cite{SM}.
To avoid finite-$N$ effects, we can work at a slightly larger $\beta$, e.g. $\beta = 0.411\pi$, whose superradiance potential is shown in Fig.~\ref{fig3}(a) and as a pink line in Fig.~\ref{fig2}(a). The potential has a dark state at $(\theta_0, \beta) = (2.41\pi, 0.411\pi)$, which is close to the critical point at $\theta_c = 2.45\pi$. Preparing the system at this value of $(\theta_0,\beta)$ will yield $\xi^2 \sim 0.05$ without any driving. This squeezing gets better as the critical point $(\theta_{c,\rm dark}, \beta_{c,\rm dark}) \approx (2.45\pi, 0.41\pi)$ is approached.

\emph{Transferring the squeezing to the dark state.---}
Scalable squeezing can also be generated with $\Omega\neq0$ in a bright state close to other critical points. This squeezing can still be transferred to a dark state, $\theta_{\rm dark}$, where it will be immune to superradiance after switching off the drive. This idea works regardless of how the squeezing was generated as long as the HP approximation is valid.

As an example, we consider spin squeezing generated at $(\theta_0, \beta) = (4.46\pi, 0.5\pi)$ [green circle in Fig.~\ref{fig3}(b)], where $\theta_0$ is close to a critical point, $\theta_c = 4.47\pi$ [Fig.~\ref{fig3}(b) and gray cut in Fig.~\ref{fig2}(a)]. 
The basic idea, after acquiring squeezing in a bright state, is to first rotate the Bloch vector to a mean-field dark state and switch off the continuous drive $\Omega$. 
This can be accomplished by a rotation $\exp[ i(\theta_0-\theta_{\rm dark})\hat D^x_{+1}]$ to the dark state at $\theta_{\rm dark} = 3.87\pi$. 
Then, to avoid losing the squeezing due to further evolution towards the new steady state at $\theta_{\rm dark}$ and $\Omega=0$, one needs to perform additional single-particle rotations that transfer the squeezed quadratures to the conserved quadratures $\hat X^b_2$, $\hat X^b_3$, $\hat Y^b_2$, or $\hat Y^b_3$. Explicit forms of these rotations are given in~\cite{SM}. Single-particle rotations for multilevel atoms can be implemented using, e.g., quantum optimal control~\cite{omanakuttan2021quantum}.
Note that all rotations should be fast compared to $N\Gamma$.

We numerically simulate this protocol using a cumulant expansion~\cite{SM} in 
Fig.~\ref{fig3}(c), which shows the evolution of the noise (variances of spin variables) for $(\theta_0, \beta) = (4.46\pi, 0.5\pi)$. At $t = (1000/N\Gamma)$ we implement the above rotations and let the system evolve freely with $\Omega=0$.
Clearly, the two most squeezed quadratures are preserved.

\emph{Best squeezing achievable.---} The best squeezing achieved is limited by \Rev{higher-order terms in the HP approximation, and single-particle decoherence set fundamentally by the single-particle linewidth $\gamma$}. In Fig.~\ref{fig3}(d), we use the cumulant expansion, which includes the higher-order HP corrections, to find the best squeezing in the two modes for the two above protocols: squeezing at a dark state or a bright state. Specifically, we scan $\theta_0$ between $2\pi$ and $2.45\pi$ along the dark manifold [dashed black line in Fig.~\ref{fig2}(b)], and $\theta_0$ between $3.87\pi$ and $4.47\pi$ at $\beta = 0.5\pi$ [Fig.~\ref{fig3}(b)].
We numerically find in all cases that the squeezing $\xi^2$ scales as $N^{-0.25}$\Rev{, and derive this scaling in ~\cite{SM}}. \Rev{We show in~\cite{SM} that free-space spontaneous emission yields squeezing which scales with $N$ as $1/\sqrt{NC}$ where $C=\Gamma/\gamma$ is the cavity-cooperativity. For $NC^2 \gg 1$, finite-size effects limit the squeezing more than spontaneous emission. 
Our protocol has the advantage that squeezing is generated in more than one quadrature and is thus useful for more general metrological tasks~\cite{pezze2018quantum}.}

\textit{Outlook.---} 
While we have focused on the case with only one relevant polarization, the ideas presented can be generalized to situations with two relevant polarizations, where up to four quadratures can be squeezed~\cite{SM}. The presented ideas open unique opportunities for the generation and storage of squeezing in generic multilevel systems even beyond the cavity setting discussed here.

\begin{acknowledgments}
We thank Jeremy T. Young, Dylan Young, and James K. Thompson for valuable discussions and carefully reading the manuscript. This work is supported by the AFOSR Grants No. FA9550-18-1-0319 and No. FA9550-19-1-0275, by the DARPA and ARO Grant No. W911NF-16-1-0576, by the NSF JILA-PFC PHY-2317149, the VBFF fellowship, QLCI-OMA-2016244, by the U.S. Department of Energy, Office of Science, National Quantum Information Science Research Centers Quantum Systems Accelerator, and by the National Institute of Standards and Technology.
\end{acknowledgments}
\bibliography{bibliography}

\begin{thebibliography}{61}%
\makeatletter
\providecommand \@ifxundefined [1]{%
 \@ifx{#1\undefined}
}%
\providecommand \@ifnum [1]{%
 \ifnum #1\expandafter \@firstoftwo
 \else \expandafter \@secondoftwo
 \fi
}%
\providecommand \@ifx [1]{%
 \ifx #1\expandafter \@firstoftwo
 \else \expandafter \@secondoftwo
 \fi
}%
\providecommand \natexlab [1]{#1}%
\providecommand \enquote  [1]{``#1''}%
\providecommand \bibnamefont  [1]{#1}%
\providecommand \bibfnamefont [1]{#1}%
\providecommand \citenamefont [1]{#1}%
\providecommand \href@noop [0]{\@secondoftwo}%
\providecommand \href [0]{\begingroup \@sanitize@url \@href}%
\providecommand \@href[1]{\@@startlink{#1}\@@href}%
\providecommand \@@href[1]{\endgroup#1\@@endlink}%
\providecommand \@sanitize@url [0]{\catcode `\\12\catcode `\$12\catcode
  `\&12\catcode `\#12\catcode `\^12\catcode `\_12\catcode `\%12\relax}%
\providecommand \@@startlink[1]{}%
\providecommand \@@endlink[0]{}%
\providecommand \url  [0]{\begingroup\@sanitize@url \@url }%
\providecommand \@url [1]{\endgroup\@href {#1}{\urlprefix }}%
\providecommand \urlprefix  [0]{URL }%
\providecommand \Eprint [0]{\href }%
\providecommand \doibase [0]{https://doi.org/}%
\providecommand \selectlanguage [0]{\@gobble}%
\providecommand \bibinfo  [0]{\@secondoftwo}%
\providecommand \bibfield  [0]{\@secondoftwo}%
\providecommand \translation [1]{[#1]}%
\providecommand \BibitemOpen [0]{}%
\providecommand \bibitemStop [0]{}%
\providecommand \bibitemNoStop [0]{.\EOS\space}%
\providecommand \EOS [0]{\spacefactor3000\relax}%
\providecommand \BibitemShut  [1]{\csname bibitem#1\endcsname}%
\let\auto@bib@innerbib\@empty
%</preamble>
\bibitem [{\citenamefont {Dicke}(1954)}]{dicke1954coherence}%
  \BibitemOpen
  \bibfield  {author} {\bibinfo {author} {\bibfnamefont {R.~H.}\ \bibnamefont
  {Dicke}},\ }\bibfield  {title} {\bibinfo {title} {Coherence in spontaneous
  radiation processes},\ }\href@noop {} {\bibfield  {journal} {\bibinfo
  {journal} {Phys. Rev.}\ }\textbf {\bibinfo {volume} {93}},\ \bibinfo {pages}
  {99} (\bibinfo {year} {1954})}\BibitemShut {NoStop}%
\bibitem [{\citenamefont {Asenjo-Garcia}\ \emph {et~al.}(2017)\citenamefont
  {Asenjo-Garcia}, \citenamefont {Moreno-Cardoner}, \citenamefont {Albrecht},
  \citenamefont {Kimble},\ and\ \citenamefont {Chang}}]{asenjo2017exponential}%
  \BibitemOpen
  \bibfield  {author} {\bibinfo {author} {\bibfnamefont {A.}~\bibnamefont
  {Asenjo-Garcia}}, \bibinfo {author} {\bibfnamefont {M.}~\bibnamefont
  {Moreno-Cardoner}}, \bibinfo {author} {\bibfnamefont {A.}~\bibnamefont
  {Albrecht}}, \bibinfo {author} {\bibfnamefont {H.~J.}\ \bibnamefont
  {Kimble}},\ and\ \bibinfo {author} {\bibfnamefont {D.~E.}\ \bibnamefont
  {Chang}},\ }\bibfield  {title} {\bibinfo {title} {Exponential improvement in
  photon storage fidelities using subradiance and ``selective radiance'' in
  atomic arrays},\ }\href@noop {} {\bibfield  {journal} {\bibinfo  {journal}
  {Phys. Rev. X}\ }\textbf {\bibinfo {volume} {7}},\ \bibinfo {pages} {031024}
  (\bibinfo {year} {2017})}\BibitemShut {NoStop}%
\bibitem [{\citenamefont {Shahmoon}\ \emph {et~al.}(2017)\citenamefont
  {Shahmoon}, \citenamefont {Wild}, \citenamefont {Lukin},\ and\ \citenamefont
  {Yelin}}]{shahmoon2017cooperative}%
  \BibitemOpen
  \bibfield  {author} {\bibinfo {author} {\bibfnamefont {E.}~\bibnamefont
  {Shahmoon}}, \bibinfo {author} {\bibfnamefont {D.~S.}\ \bibnamefont {Wild}},
  \bibinfo {author} {\bibfnamefont {M.~D.}\ \bibnamefont {Lukin}},\ and\
  \bibinfo {author} {\bibfnamefont {S.~F.}\ \bibnamefont {Yelin}},\ }\bibfield
  {title} {\bibinfo {title} {Cooperative resonances in light scattering from
  two-dimensional atomic arrays},\ }\href@noop {} {\bibfield  {journal}
  {\bibinfo  {journal} {Phys. Rev. Lett.}\ }\textbf {\bibinfo {volume} {118}},\
  \bibinfo {pages} {113601} (\bibinfo {year} {2017})}\BibitemShut {NoStop}%
\bibitem [{\citenamefont {Rui}\ \emph {et~al.}(2020)\citenamefont {Rui},
  \citenamefont {Wei}, \citenamefont {Rubio-Abadal}, \citenamefont {Hollerith},
  \citenamefont {Zeiher}, \citenamefont {Stamper-Kurn}, \citenamefont {Gross},\
  and\ \citenamefont {Bloch}}]{Rui2020}%
  \BibitemOpen
  \bibfield  {author} {\bibinfo {author} {\bibfnamefont {J.}~\bibnamefont
  {Rui}}, \bibinfo {author} {\bibfnamefont {D.}~\bibnamefont {Wei}}, \bibinfo
  {author} {\bibfnamefont {A.}~\bibnamefont {Rubio-Abadal}}, \bibinfo {author}
  {\bibfnamefont {S.}~\bibnamefont {Hollerith}}, \bibinfo {author}
  {\bibfnamefont {J.}~\bibnamefont {Zeiher}}, \bibinfo {author} {\bibfnamefont
  {D.~M.}\ \bibnamefont {Stamper-Kurn}}, \bibinfo {author} {\bibfnamefont
  {C.}~\bibnamefont {Gross}},\ and\ \bibinfo {author} {\bibfnamefont
  {I.}~\bibnamefont {Bloch}},\ }\bibfield  {title} {\bibinfo {title} {A
  subradiant optical mirror formed by a single structured atomic layer},\
  }\href {https://doi.org/10.1038/s41586-020-2463-x} {\bibfield  {journal}
  {\bibinfo  {journal} {Nature (London)}\ }\textbf {\bibinfo {volume} {583}},\
  \bibinfo {pages} {369} (\bibinfo {year} {2020})}\BibitemShut {NoStop}%
\bibitem [{\citenamefont {Needham}\ \emph {et~al.}(2019)\citenamefont
  {Needham}, \citenamefont {Lesanovsky},\ and\ \citenamefont
  {Olmos}}]{needham2019subradiance}%
  \BibitemOpen
  \bibfield  {author} {\bibinfo {author} {\bibfnamefont {J.~A.}\ \bibnamefont
  {Needham}}, \bibinfo {author} {\bibfnamefont {I.}~\bibnamefont
  {Lesanovsky}},\ and\ \bibinfo {author} {\bibfnamefont {B.}~\bibnamefont
  {Olmos}},\ }\bibfield  {title} {\bibinfo {title} {Subradiance-protected
  excitation transport},\ }\href@noop {} {\bibfield  {journal} {\bibinfo
  {journal} {New J. Phys.}\ }\textbf {\bibinfo {volume} {21}},\ \bibinfo
  {pages} {073061} (\bibinfo {year} {2019})}\BibitemShut {NoStop}%
\bibitem [{\citenamefont {Ballantine}\ and\ \citenamefont
  {Ruostekoski}(2020)}]{ballantine2020subradiance}%
  \BibitemOpen
  \bibfield  {author} {\bibinfo {author} {\bibfnamefont {K.~E.}\ \bibnamefont
  {Ballantine}}\ and\ \bibinfo {author} {\bibfnamefont {J.}~\bibnamefont
  {Ruostekoski}},\ }\bibfield  {title} {\bibinfo {title} {Subradiance-protected
  excitation spreading in the generation of collimated photon emission from an
  atomic array},\ }\href@noop {} {\bibfield  {journal} {\bibinfo  {journal}
  {Phys. Rev. Res.}\ }\textbf {\bibinfo {volume} {2}},\ \bibinfo {pages}
  {023086} (\bibinfo {year} {2020})}\BibitemShut {NoStop}%
\bibitem [{\citenamefont {Perczel}\ \emph
  {et~al.}(2017{\natexlab{a}})\citenamefont {Perczel}, \citenamefont
  {Borregaard}, \citenamefont {Chang}, \citenamefont {Pichler}, \citenamefont
  {Yelin}, \citenamefont {Zoller},\ and\ \citenamefont
  {Lukin}}]{perczel2017topological}%
  \BibitemOpen
  \bibfield  {author} {\bibinfo {author} {\bibfnamefont {J.}~\bibnamefont
  {Perczel}}, \bibinfo {author} {\bibfnamefont {J.}~\bibnamefont {Borregaard}},
  \bibinfo {author} {\bibfnamefont {D.~E.}\ \bibnamefont {Chang}}, \bibinfo
  {author} {\bibfnamefont {H.}~\bibnamefont {Pichler}}, \bibinfo {author}
  {\bibfnamefont {S.~F.}\ \bibnamefont {Yelin}}, \bibinfo {author}
  {\bibfnamefont {P.}~\bibnamefont {Zoller}},\ and\ \bibinfo {author}
  {\bibfnamefont {M.~D.}\ \bibnamefont {Lukin}},\ }\bibfield  {title} {\bibinfo
  {title} {Topological quantum optics in two-dimensional atomic arrays},\
  }\href@noop {} {\bibfield  {journal} {\bibinfo  {journal} {Phys. Rev. Lett.}\
  }\textbf {\bibinfo {volume} {119}},\ \bibinfo {pages} {023603} (\bibinfo
  {year} {2017}{\natexlab{a}})}\BibitemShut {NoStop}%
\bibitem [{\citenamefont {Perczel}\ \emph
  {et~al.}(2017{\natexlab{b}})\citenamefont {Perczel}, \citenamefont
  {Borregaard}, \citenamefont {Chang}, \citenamefont {Pichler}, \citenamefont
  {Yelin}, \citenamefont {Zoller},\ and\ \citenamefont
  {Lukin}}]{perczel2017photonic}%
  \BibitemOpen
  \bibfield  {author} {\bibinfo {author} {\bibfnamefont {J.}~\bibnamefont
  {Perczel}}, \bibinfo {author} {\bibfnamefont {J.}~\bibnamefont {Borregaard}},
  \bibinfo {author} {\bibfnamefont {D.~E.}\ \bibnamefont {Chang}}, \bibinfo
  {author} {\bibfnamefont {H.}~\bibnamefont {Pichler}}, \bibinfo {author}
  {\bibfnamefont {S.~F.}\ \bibnamefont {Yelin}}, \bibinfo {author}
  {\bibfnamefont {P.}~\bibnamefont {Zoller}},\ and\ \bibinfo {author}
  {\bibfnamefont {M.~D.}\ \bibnamefont {Lukin}},\ }\bibfield  {title} {\bibinfo
  {title} {Photonic band structure of two-dimensional atomic lattices},\
  }\href@noop {} {\bibfield  {journal} {\bibinfo  {journal} {Phys. Rev. A}\
  }\textbf {\bibinfo {volume} {96}},\ \bibinfo {pages} {063801} (\bibinfo
  {year} {2017}{\natexlab{b}})}\BibitemShut {NoStop}%
\bibitem [{\citenamefont {Bettles}\ \emph {et~al.}(2017)\citenamefont
  {Bettles}, \citenamefont {Minar}, \citenamefont {Adams}, \citenamefont
  {Lesanovsky},\ and\ \citenamefont {Olmos}}]{bettles2017topological}%
  \BibitemOpen
  \bibfield  {author} {\bibinfo {author} {\bibfnamefont {R.~J.}\ \bibnamefont
  {Bettles}}, \bibinfo {author} {\bibfnamefont {J.}~\bibnamefont {Minar}},
  \bibinfo {author} {\bibfnamefont {C.~S.}\ \bibnamefont {Adams}}, \bibinfo
  {author} {\bibfnamefont {I.}~\bibnamefont {Lesanovsky}},\ and\ \bibinfo
  {author} {\bibfnamefont {B.}~\bibnamefont {Olmos}},\ }\bibfield  {title}
  {\bibinfo {title} {Topological properties of a dense atomic lattice gas},\
  }\href@noop {} {\bibfield  {journal} {\bibinfo  {journal} {Phys. Rev. A}\
  }\textbf {\bibinfo {volume} {96}},\ \bibinfo {pages} {041603(R)} (\bibinfo
  {year} {2017})}\BibitemShut {NoStop}%
\bibitem [{\citenamefont {Zhang}\ \emph {et~al.}(2019)\citenamefont {Zhang},
  \citenamefont {Chen}, \citenamefont {Yakovlev},\ and\ \citenamefont
  {Yuan}}]{zhang2019tunable}%
  \BibitemOpen
  \bibfield  {author} {\bibinfo {author} {\bibfnamefont {A.}~\bibnamefont
  {Zhang}}, \bibinfo {author} {\bibfnamefont {X.}~\bibnamefont {Chen}},
  \bibinfo {author} {\bibfnamefont {V.~V.}\ \bibnamefont {Yakovlev}},\ and\
  \bibinfo {author} {\bibfnamefont {L.}~\bibnamefont {Yuan}},\ }\bibfield
  {title} {\bibinfo {title} {Tunable topologically-protected super-and
  subradiant boundary states in one-dimensional atomic arrays},\ }\href@noop {}
  {\bibfield  {journal} {\bibinfo  {journal} {Commun. Phys.}\ }\textbf
  {\bibinfo {volume} {2}},\ \bibinfo {pages} {157} (\bibinfo {year}
  {2019})}\BibitemShut {NoStop}%
\bibitem [{\citenamefont {Syzranov}\ \emph {et~al.}(2016)\citenamefont
  {Syzranov}, \citenamefont {Wall}, \citenamefont {Zhu}, \citenamefont
  {Gurarie},\ and\ \citenamefont {Rey}}]{syzranov2016emergent}%
  \BibitemOpen
  \bibfield  {author} {\bibinfo {author} {\bibfnamefont {S.~V.}\ \bibnamefont
  {Syzranov}}, \bibinfo {author} {\bibfnamefont {M.~L.}\ \bibnamefont {Wall}},
  \bibinfo {author} {\bibfnamefont {B.}~\bibnamefont {Zhu}}, \bibinfo {author}
  {\bibfnamefont {V.}~\bibnamefont {Gurarie}},\ and\ \bibinfo {author}
  {\bibfnamefont {A.~M.}\ \bibnamefont {Rey}},\ }\bibfield  {title} {\bibinfo
  {title} {Emergent weyl excitations in systems of polar particles},\
  }\href@noop {} {\bibfield  {journal} {\bibinfo  {journal} {Nat. Commun.}\
  }\textbf {\bibinfo {volume} {7}},\ \bibinfo {pages} {1} (\bibinfo {year}
  {2016})}\BibitemShut {NoStop}%
\bibitem [{\citenamefont {Gonz{\'a}lez-Tudela}\ \emph
  {et~al.}(2015)\citenamefont {Gonz{\'a}lez-Tudela}, \citenamefont {Paulisch},
  \citenamefont {Chang}, \citenamefont {Kimble},\ and\ \citenamefont
  {Cirac}}]{gonzalez2015deterministic}%
  \BibitemOpen
  \bibfield  {author} {\bibinfo {author} {\bibfnamefont {A.}~\bibnamefont
  {Gonz{\'a}lez-Tudela}}, \bibinfo {author} {\bibfnamefont {V.}~\bibnamefont
  {Paulisch}}, \bibinfo {author} {\bibfnamefont {D.~E.}\ \bibnamefont {Chang}},
  \bibinfo {author} {\bibfnamefont {H.~J.}\ \bibnamefont {Kimble}},\ and\
  \bibinfo {author} {\bibfnamefont {J.~I.}\ \bibnamefont {Cirac}},\ }\bibfield
  {title} {\bibinfo {title} {Deterministic generation of arbitrary photonic
  states assisted by dissipation},\ }\href@noop {} {\bibfield  {journal}
  {\bibinfo  {journal} {Phys. Rev. Lett.}\ }\textbf {\bibinfo {volume} {115}},\
  \bibinfo {pages} {163603} (\bibinfo {year} {2015})}\BibitemShut {NoStop}%
\bibitem [{\citenamefont {Henriet}\ \emph {et~al.}(2019)\citenamefont
  {Henriet}, \citenamefont {Douglas}, \citenamefont {Chang},\ and\
  \citenamefont {Albrecht}}]{henriet2019critical}%
  \BibitemOpen
  \bibfield  {author} {\bibinfo {author} {\bibfnamefont {L.}~\bibnamefont
  {Henriet}}, \bibinfo {author} {\bibfnamefont {J.~S.}\ \bibnamefont
  {Douglas}}, \bibinfo {author} {\bibfnamefont {D.~E.}\ \bibnamefont {Chang}},\
  and\ \bibinfo {author} {\bibfnamefont {A.}~\bibnamefont {Albrecht}},\
  }\bibfield  {title} {\bibinfo {title} {Critical open-system dynamics in a
  one-dimensional optical-lattice clock},\ }\href@noop {} {\bibfield  {journal}
  {\bibinfo  {journal} {Phys. Rev. A}\ }\textbf {\bibinfo {volume} {99}},\
  \bibinfo {pages} {023802} (\bibinfo {year} {2019})}\BibitemShut {NoStop}%
\bibitem [{\citenamefont {Qu}\ and\ \citenamefont {Rey}(2019)}]{Chu2019}%
  \BibitemOpen
  \bibfield  {author} {\bibinfo {author} {\bibfnamefont {C.}~\bibnamefont
  {Qu}}\ and\ \bibinfo {author} {\bibfnamefont {A.~M.}\ \bibnamefont {Rey}},\
  }\bibfield  {title} {\bibinfo {title} {Spin squeezing and many-body dipolar
  dynamics in optical lattice clocks},\ }\href
  {https://doi.org/10.1103/PhysRevA.100.041602} {\bibfield  {journal} {\bibinfo
   {journal} {Phys. Rev. A}\ }\textbf {\bibinfo {volume} {100}},\ \bibinfo
  {pages} {041602(R)} (\bibinfo {year} {2019})}\BibitemShut {NoStop}%
\bibitem [{\citenamefont {Pi\~neiro Orioli}\ and\ \citenamefont
  {Rey}(2019)}]{Pineiro2019}%
  \BibitemOpen
  \bibfield  {author} {\bibinfo {author} {\bibfnamefont {A.}~\bibnamefont
  {Pi\~neiro Orioli}}\ and\ \bibinfo {author} {\bibfnamefont {A.~M.}\
  \bibnamefont {Rey}},\ }\bibfield  {title} {\bibinfo {title} {Dark states of
  multilevel fermionic atoms in doubly filled optical lattices},\ }\href
  {https://doi.org/10.1103/PhysRevLett.123.223601} {\bibfield  {journal}
  {\bibinfo  {journal} {Phys. Rev. Lett.}\ }\textbf {\bibinfo {volume} {123}},\
  \bibinfo {pages} {223601} (\bibinfo {year} {2019})}\BibitemShut {NoStop}%
\bibitem [{\citenamefont {Pi\~neiro Orioli}\ and\ \citenamefont
  {Rey}(2020)}]{Pineiro2020}%
  \BibitemOpen
  \bibfield  {author} {\bibinfo {author} {\bibfnamefont {A.}~\bibnamefont
  {Pi\~neiro Orioli}}\ and\ \bibinfo {author} {\bibfnamefont {A.~M.}\
  \bibnamefont {Rey}},\ }\bibfield  {title} {\bibinfo {title} {Subradiance of
  multilevel fermionic atoms in arrays with filling $n\ensuremath{\ge}2$},\
  }\href {https://doi.org/10.1103/PhysRevA.101.043816} {\bibfield  {journal}
  {\bibinfo  {journal} {Phys. Rev. A}\ }\textbf {\bibinfo {volume} {101}},\
  \bibinfo {pages} {043816} (\bibinfo {year} {2020})}\BibitemShut {NoStop}%
\bibitem [{\citenamefont {Ma}\ \emph {et~al.}(2011)\citenamefont {Ma},
  \citenamefont {Wang}, \citenamefont {Sun},\ and\ \citenamefont
  {Nori}}]{ma2011quantum}%
  \BibitemOpen
  \bibfield  {author} {\bibinfo {author} {\bibfnamefont {J.}~\bibnamefont
  {Ma}}, \bibinfo {author} {\bibfnamefont {X.}~\bibnamefont {Wang}}, \bibinfo
  {author} {\bibfnamefont {C.-P.}\ \bibnamefont {Sun}},\ and\ \bibinfo {author}
  {\bibfnamefont {F.}~\bibnamefont {Nori}},\ }\bibfield  {title} {\bibinfo
  {title} {Quantum spin squeezing},\ }\href@noop {} {\bibfield  {journal}
  {\bibinfo  {journal} {Phys. Rep.}\ }\textbf {\bibinfo {volume} {509}},\
  \bibinfo {pages} {89} (\bibinfo {year} {2011})}\BibitemShut {NoStop}%
\bibitem [{\citenamefont {Schleier-Smith}\ \emph {et~al.}(2010)\citenamefont
  {Schleier-Smith}, \citenamefont {Leroux},\ and\ \citenamefont
  {Vuleti{\'c}}}]{schleier2010squeezing}%
  \BibitemOpen
  \bibfield  {author} {\bibinfo {author} {\bibfnamefont {M.~H.}\ \bibnamefont
  {Schleier-Smith}}, \bibinfo {author} {\bibfnamefont {I.~D.}\ \bibnamefont
  {Leroux}},\ and\ \bibinfo {author} {\bibfnamefont {V.}~\bibnamefont
  {Vuleti{\'c}}},\ }\bibfield  {title} {\bibinfo {title} {Squeezing the
  collective spin of a dilute atomic ensemble by cavity feedback},\ }\href@noop
  {} {\bibfield  {journal} {\bibinfo  {journal} {Phys. Rev. A}\ }\textbf
  {\bibinfo {volume} {81}},\ \bibinfo {pages} {021804(R)} (\bibinfo {year}
  {2010})}\BibitemShut {NoStop}%
\bibitem [{\citenamefont {Leroux}\ \emph {et~al.}(2010)\citenamefont {Leroux},
  \citenamefont {Schleier-Smith},\ and\ \citenamefont
  {Vuleti{\'c}}}]{leroux2010implementation}%
  \BibitemOpen
  \bibfield  {author} {\bibinfo {author} {\bibfnamefont {I.~D.}\ \bibnamefont
  {Leroux}}, \bibinfo {author} {\bibfnamefont {M.~H.}\ \bibnamefont
  {Schleier-Smith}},\ and\ \bibinfo {author} {\bibfnamefont {V.}~\bibnamefont
  {Vuleti{\'c}}},\ }\bibfield  {title} {\bibinfo {title} {Implementation of
  cavity squeezing of a collective atomic spin},\ }\href@noop {} {\bibfield
  {journal} {\bibinfo  {journal} {Phys. Rev. Lett.}\ }\textbf {\bibinfo
  {volume} {104}},\ \bibinfo {pages} {073602} (\bibinfo {year}
  {2010})}\BibitemShut {NoStop}%
\bibitem [{\citenamefont {Chen}\ \emph {et~al.}(2014)\citenamefont {Chen},
  \citenamefont {Bohnet}, \citenamefont {Weiner}, \citenamefont {Cox},\ and\
  \citenamefont {Thompson}}]{chen2014cavity}%
  \BibitemOpen
  \bibfield  {author} {\bibinfo {author} {\bibfnamefont {Z.}~\bibnamefont
  {Chen}}, \bibinfo {author} {\bibfnamefont {J.~G.}\ \bibnamefont {Bohnet}},
  \bibinfo {author} {\bibfnamefont {J.~M.}\ \bibnamefont {Weiner}}, \bibinfo
  {author} {\bibfnamefont {K.~C.}\ \bibnamefont {Cox}},\ and\ \bibinfo {author}
  {\bibfnamefont {J.~K.}\ \bibnamefont {Thompson}},\ }\bibfield  {title}
  {\bibinfo {title} {Cavity-aided nondemolition measurements for atom counting
  and spin squeezing},\ }\href@noop {} {\bibfield  {journal} {\bibinfo
  {journal} {Phys. Rev. A}\ }\textbf {\bibinfo {volume} {89}},\ \bibinfo
  {pages} {043837} (\bibinfo {year} {2014})}\BibitemShut {NoStop}%
\bibitem [{\citenamefont {Cox}\ \emph {et~al.}(2016)\citenamefont {Cox},
  \citenamefont {Greve}, \citenamefont {Weiner},\ and\ \citenamefont
  {Thompson}}]{Cox2016}%
  \BibitemOpen
  \bibfield  {author} {\bibinfo {author} {\bibfnamefont {K.~C.}\ \bibnamefont
  {Cox}}, \bibinfo {author} {\bibfnamefont {G.~P.}\ \bibnamefont {Greve}},
  \bibinfo {author} {\bibfnamefont {J.~M.}\ \bibnamefont {Weiner}},\ and\
  \bibinfo {author} {\bibfnamefont {J.~K.}\ \bibnamefont {Thompson}},\
  }\bibfield  {title} {\bibinfo {title} {Deterministic squeezed states with
  collective measurements and feedback},\ }\href
  {https://doi.org/10.1103/PhysRevLett.116.093602} {\bibfield  {journal}
  {\bibinfo  {journal} {Phys. Rev. Lett.}\ }\textbf {\bibinfo {volume} {116}},\
  \bibinfo {pages} {093602} (\bibinfo {year} {2016})}\BibitemShut {NoStop}%
\bibitem [{\citenamefont {Hosten}\ \emph {et~al.}(2016)\citenamefont {Hosten},
  \citenamefont {Engelsen}, \citenamefont {Krishnakumar},\ and\ \citenamefont
  {Kasevich}}]{hosten2016measurement}%
  \BibitemOpen
  \bibfield  {author} {\bibinfo {author} {\bibfnamefont {O.}~\bibnamefont
  {Hosten}}, \bibinfo {author} {\bibfnamefont {N.~J.}\ \bibnamefont
  {Engelsen}}, \bibinfo {author} {\bibfnamefont {R.}~\bibnamefont
  {Krishnakumar}},\ and\ \bibinfo {author} {\bibfnamefont {M.~A.}\ \bibnamefont
  {Kasevich}},\ }\bibfield  {title} {\bibinfo {title} {Measurement noise 100
  times lower than the quantum-projection limit using entangled atoms},\
  }\href@noop {} {\bibfield  {journal} {\bibinfo  {journal} {Nature (London)}\
  }\textbf {\bibinfo {volume} {529}},\ \bibinfo {pages} {505} (\bibinfo {year}
  {2016})}\BibitemShut {NoStop}%
\bibitem [{\citenamefont {Pezz\`e}\ \emph {et~al.}(2018)\citenamefont
  {Pezz\`e}, \citenamefont {Smerzi}, \citenamefont {Oberthaler}, \citenamefont
  {Schmied},\ and\ \citenamefont {Treutlein}}]{Pezze2018}%
  \BibitemOpen
  \bibfield  {author} {\bibinfo {author} {\bibfnamefont {L.}~\bibnamefont
  {Pezz\`e}}, \bibinfo {author} {\bibfnamefont {A.}~\bibnamefont {Smerzi}},
  \bibinfo {author} {\bibfnamefont {M.~K.}\ \bibnamefont {Oberthaler}},
  \bibinfo {author} {\bibfnamefont {R.}~\bibnamefont {Schmied}},\ and\ \bibinfo
  {author} {\bibfnamefont {P.}~\bibnamefont {Treutlein}},\ }\bibfield  {title}
  {\bibinfo {title} {Quantum metrology with nonclassical states of atomic
  ensembles},\ }\href {https://doi.org/10.1103/RevModPhys.90.035005} {\bibfield
   {journal} {\bibinfo  {journal} {Rev. Mod. Phys.}\ }\textbf {\bibinfo
  {volume} {90}},\ \bibinfo {pages} {035005} (\bibinfo {year}
  {2018})}\BibitemShut {NoStop}%
\bibitem [{\citenamefont {Gross}\ and\ \citenamefont
  {Haroche}(1982)}]{GrossHarocheSuperr}%
  \BibitemOpen
  \bibfield  {author} {\bibinfo {author} {\bibfnamefont {M.}~\bibnamefont
  {Gross}}\ and\ \bibinfo {author} {\bibfnamefont {S.}~\bibnamefont
  {Haroche}},\ }\bibfield  {title} {\bibinfo {title} {Superradiance: An essay
  on the theory of collective spontaneous emission},\ }\href
  {https://doi.org/https://doi.org/10.1016/0370-1573(82)90102-8} {\bibfield
  {journal} {\bibinfo  {journal} {Phys. Rep.}\ }\textbf {\bibinfo {volume}
  {93}},\ \bibinfo {pages} {301 } (\bibinfo {year} {1982})}\BibitemShut
  {NoStop}%
\bibitem [{\citenamefont {Norcia}\ \emph {et~al.}(2016)\citenamefont {Norcia},
  \citenamefont {Winchester}, \citenamefont {Cline},\ and\ \citenamefont
  {Thompson}}]{norcia2016superradiance}%
  \BibitemOpen
  \bibfield  {author} {\bibinfo {author} {\bibfnamefont {M.~A.}\ \bibnamefont
  {Norcia}}, \bibinfo {author} {\bibfnamefont {M.~N.}\ \bibnamefont
  {Winchester}}, \bibinfo {author} {\bibfnamefont {J.~R.~K.}\ \bibnamefont
  {Cline}},\ and\ \bibinfo {author} {\bibfnamefont {J.~K.}\ \bibnamefont
  {Thompson}},\ }\bibfield  {title} {\bibinfo {title} {Superradiance on the
  millihertz linewidth strontium clock transition},\ }\href@noop {} {\bibfield
  {journal} {\bibinfo  {journal} {Sci. Adv.}\ }\textbf {\bibinfo {volume}
  {2}},\ \bibinfo {pages} {e1601231} (\bibinfo {year} {2016})}\BibitemShut
  {NoStop}%
\bibitem [{\citenamefont {Barberena}\ \emph {et~al.}(2019)\citenamefont
  {Barberena}, \citenamefont {Lewis-Swan}, \citenamefont {Thompson},\ and\
  \citenamefont {Rey}}]{barberena2019driven}%
  \BibitemOpen
  \bibfield  {author} {\bibinfo {author} {\bibfnamefont {D.}~\bibnamefont
  {Barberena}}, \bibinfo {author} {\bibfnamefont {R.~J.}\ \bibnamefont
  {Lewis-Swan}}, \bibinfo {author} {\bibfnamefont {J.~K.}\ \bibnamefont
  {Thompson}},\ and\ \bibinfo {author} {\bibfnamefont {A.~M.}\ \bibnamefont
  {Rey}},\ }\bibfield  {title} {\bibinfo {title} {Driven-dissipative quantum
  dynamics in ultra-long-lived dipoles in an optical cavity},\ }\href@noop {}
  {\bibfield  {journal} {\bibinfo  {journal} {Phys. Rev. A}\ }\textbf {\bibinfo
  {volume} {99}},\ \bibinfo {pages} {053411} (\bibinfo {year}
  {2019})}\BibitemShut {NoStop}%
\bibitem [{\citenamefont {Wolfe}\ and\ \citenamefont
  {Yelin}()}]{wolfe2014spin}%
  \BibitemOpen
  \bibfield  {author} {\bibinfo {author} {\bibfnamefont {E.}~\bibnamefont
  {Wolfe}}\ and\ \bibinfo {author} {\bibfnamefont {S.~F.}\ \bibnamefont
  {Yelin}},\ }\bibfield  {title} {\bibinfo {title} {Spin squeezing by means of
  driven superradiance},\ }\href@noop {} {\bibinfo  {journal}
  {arXiv:1405.5288}\ }\BibitemShut {NoStop}%
\bibitem [{\citenamefont {Dalla~Torre}\ \emph {et~al.}(2013)\citenamefont
  {Dalla~Torre}, \citenamefont {Otterbach}, \citenamefont {Demler},
  \citenamefont {Vuletic},\ and\ \citenamefont {Lukin}}]{dalla2013dissipative}%
  \BibitemOpen
\bibfield  {journal} {  }\bibfield  {author} {\bibinfo {author} {\bibfnamefont
  {E.~G.}\ \bibnamefont {Dalla~Torre}}, \bibinfo {author} {\bibfnamefont
  {J.}~\bibnamefont {Otterbach}}, \bibinfo {author} {\bibfnamefont
  {E.}~\bibnamefont {Demler}}, \bibinfo {author} {\bibfnamefont
  {V.}~\bibnamefont {Vuletic}},\ and\ \bibinfo {author} {\bibfnamefont {M.~D.}\
  \bibnamefont {Lukin}},\ }\bibfield  {title} {\bibinfo {title} {Dissipative
  preparation of spin squeezed atomic ensembles in a steady state},\
  }\href@noop {} {\bibfield  {journal} {\bibinfo  {journal} {Phys. Rev. Lett.}\
  }\textbf {\bibinfo {volume} {110}},\ \bibinfo {pages} {120402} (\bibinfo
  {year} {2013})}\BibitemShut {NoStop}%
\bibitem [{\citenamefont {Walls}\ \emph {et~al.}(1978)\citenamefont {Walls},
  \citenamefont {Drummond}, \citenamefont {Hassan},\ and\ \citenamefont
  {Carmichael}}]{walls1978non}%
  \BibitemOpen
  \bibfield  {author} {\bibinfo {author} {\bibfnamefont {D.~F.}\ \bibnamefont
  {Walls}}, \bibinfo {author} {\bibfnamefont {P.~D.}\ \bibnamefont {Drummond}},
  \bibinfo {author} {\bibfnamefont {S.~S.}\ \bibnamefont {Hassan}},\ and\
  \bibinfo {author} {\bibfnamefont {H.~J.}\ \bibnamefont {Carmichael}},\
  }\bibfield  {title} {\bibinfo {title} {Non-equilibrium phase transitions in
  cooperative atomic systems},\ }\href@noop {} {\bibfield  {journal} {\bibinfo
  {journal} {Prog. Theor. Phys. Supp.}\ }\textbf {\bibinfo {volume} {64}},\
  \bibinfo {pages} {307} (\bibinfo {year} {1978})}\BibitemShut {NoStop}%
\bibitem [{\citenamefont {Drummond}\ and\ \citenamefont
  {Carmichael}(1978)}]{drummond1978volterra}%
  \BibitemOpen
  \bibfield  {author} {\bibinfo {author} {\bibfnamefont {P.~D.}\ \bibnamefont
  {Drummond}}\ and\ \bibinfo {author} {\bibfnamefont {H.~J.}\ \bibnamefont
  {Carmichael}},\ }\bibfield  {title} {\bibinfo {title} {Volterra cycles and
  the cooperative fluorescence critical point},\ }\href@noop {} {\bibfield
  {journal} {\bibinfo  {journal} {Opt. Commun.}\ }\textbf {\bibinfo {volume}
  {27}},\ \bibinfo {pages} {160} (\bibinfo {year} {1978})}\BibitemShut
  {NoStop}%
\bibitem [{\citenamefont {Drummond}\ and\ \citenamefont
  {Hassan}(1980)}]{drummond1980multiple}%
  \BibitemOpen
  \bibfield  {author} {\bibinfo {author} {\bibfnamefont {P.~D.}\ \bibnamefont
  {Drummond}}\ and\ \bibinfo {author} {\bibfnamefont {S.~S.}\ \bibnamefont
  {Hassan}},\ }\bibfield  {title} {\bibinfo {title} {Multiple sidebands in
  cooperative resonance fluorescence: Exact semiclassical results},\
  }\href@noop {} {\bibfield  {journal} {\bibinfo  {journal} {Phys. Rev. A}\
  }\textbf {\bibinfo {volume} {22}},\ \bibinfo {pages} {662} (\bibinfo {year}
  {1980})}\BibitemShut {NoStop}%
\bibitem [{\citenamefont {Drummond}(1980)}]{drummond1980observables}%
  \BibitemOpen
  \bibfield  {author} {\bibinfo {author} {\bibfnamefont {P.~D.}\ \bibnamefont
  {Drummond}},\ }\bibfield  {title} {\bibinfo {title} {Observables and moments
  of cooperative resonance fluorescence},\ }\href@noop {} {\bibfield  {journal}
  {\bibinfo  {journal} {Phys. Rev. A}\ }\textbf {\bibinfo {volume} {22}},\
  \bibinfo {pages} {1179} (\bibinfo {year} {1980})}\BibitemShut {NoStop}%
\bibitem [{\citenamefont {Carmichael}(1980)}]{carmichael1980analytical}%
  \BibitemOpen
  \bibfield  {author} {\bibinfo {author} {\bibfnamefont {H.~J.}\ \bibnamefont
  {Carmichael}},\ }\bibfield  {title} {\bibinfo {title} {Analytical and
  numerical results for the steady state in cooperative resonance
  fluorescence},\ }\href@noop {} {\bibfield  {journal} {\bibinfo  {journal} {J.
  Phys. B (1968-1987)}\ }\textbf {\bibinfo {volume} {13}},\ \bibinfo {pages}
  {3551} (\bibinfo {year} {1980})}\BibitemShut {NoStop}%
\bibitem [{\citenamefont {Puri}\ and\ \citenamefont
  {Lawande}(1979)}]{puri1979exact}%
  \BibitemOpen
  \bibfield  {author} {\bibinfo {author} {\bibfnamefont {R.~R.}\ \bibnamefont
  {Puri}}\ and\ \bibinfo {author} {\bibfnamefont {S.~V.}\ \bibnamefont
  {Lawande}},\ }\bibfield  {title} {\bibinfo {title} {Exact steady-state
  density operator for a collective atomic system in an external field},\
  }\href@noop {} {\bibfield  {journal} {\bibinfo  {journal} {Phys. Lett. A}\
  }\textbf {\bibinfo {volume} {72}},\ \bibinfo {pages} {200} (\bibinfo {year}
  {1979})}\BibitemShut {NoStop}%
\bibitem [{\citenamefont {Somech}\ and\ \citenamefont
  {Shahmoon}()}]{somech2022quantum}%
  \BibitemOpen
  \bibfield  {author} {\bibinfo {author} {\bibfnamefont {O.}~\bibnamefont
  {Somech}}\ and\ \bibinfo {author} {\bibfnamefont {E.}~\bibnamefont
  {Shahmoon}},\ }\bibfield  {title} {\bibinfo {title} {Quantum entangled states
  of a classically radiating macroscopic spin},\ }\href@noop {} {\bibinfo
  {journal} {arXiv:2204.05455}\ }\BibitemShut {NoStop}%
\bibitem [{\citenamefont {Pi{\~n}eiro~Orioli}\ \emph
  {et~al.}(2022)\citenamefont {Pi{\~n}eiro~Orioli}, \citenamefont {Thompson},\
  and\ \citenamefont {Rey}}]{orioli2022emergent}%
  \BibitemOpen
\bibfield  {journal} {  }\bibfield  {author} {\bibinfo {author} {\bibfnamefont
  {A.}~\bibnamefont {Pi{\~n}eiro~Orioli}}, \bibinfo {author} {\bibfnamefont
  {J.~K.}\ \bibnamefont {Thompson}},\ and\ \bibinfo {author} {\bibfnamefont
  {A.~M.}\ \bibnamefont {Rey}},\ }\bibfield  {title} {\bibinfo {title}
  {Emergent dark states from superradiant dynamics in multilevel atoms in a
  cavity},\ }\href@noop {} {\bibfield  {journal} {\bibinfo  {journal} {Phys.
  Rev. X}\ }\textbf {\bibinfo {volume} {12}},\ \bibinfo {pages} {011054}
  (\bibinfo {year} {2022})}\BibitemShut {NoStop}%
\bibitem [{\citenamefont {Lin}\ \emph {et~al.}(2022)\citenamefont {Lin},
  \citenamefont {Rosa-Medina}, \citenamefont {Ferri}, \citenamefont {Finger},
  \citenamefont {Kroeger}, \citenamefont {Donner}, \citenamefont {Esslinger},\
  and\ \citenamefont {Chitra}}]{lin2022dissipation}%
  \BibitemOpen
  \bibfield  {author} {\bibinfo {author} {\bibfnamefont {R.}~\bibnamefont
  {Lin}}, \bibinfo {author} {\bibfnamefont {R.}~\bibnamefont {Rosa-Medina}},
  \bibinfo {author} {\bibfnamefont {F.}~\bibnamefont {Ferri}}, \bibinfo
  {author} {\bibfnamefont {F.}~\bibnamefont {Finger}}, \bibinfo {author}
  {\bibfnamefont {K.}~\bibnamefont {Kroeger}}, \bibinfo {author} {\bibfnamefont
  {T.}~\bibnamefont {Donner}}, \bibinfo {author} {\bibfnamefont
  {T.}~\bibnamefont {Esslinger}},\ and\ \bibinfo {author} {\bibfnamefont
  {R.}~\bibnamefont {Chitra}},\ }\bibfield  {title} {\bibinfo {title}
  {Dissipation-engineered family of nearly dark states in many-body cavity-atom
  systems},\ }\href@noop {} {\bibfield  {journal} {\bibinfo  {journal} {Phys.
  Rev. Lett.}\ }\textbf {\bibinfo {volume} {128}},\ \bibinfo {pages} {153601}
  (\bibinfo {year} {2022})}\BibitemShut {NoStop}%
\bibitem [{\citenamefont {Fan}\ and\ \citenamefont
  {Jia}(2023)}]{fan2023collective}%
  \BibitemOpen
  \bibfield  {author} {\bibinfo {author} {\bibfnamefont {J.}~\bibnamefont
  {Fan}}\ and\ \bibinfo {author} {\bibfnamefont {S.}~\bibnamefont {Jia}},\
  }\bibfield  {title} {\bibinfo {title} {Collective dynamics of the unbalanced
  three-level dicke model},\ }\href@noop {} {\bibfield  {journal} {\bibinfo
  {journal} {Phys. Rev. A}\ }\textbf {\bibinfo {volume} {107}},\ \bibinfo
  {pages} {033711} (\bibinfo {year} {2023})}\BibitemShut {NoStop}%
\bibitem [{\citenamefont {Norris}\ \emph {et~al.}(2012)\citenamefont {Norris},
  \citenamefont {Trail}, \citenamefont {Jessen},\ and\ \citenamefont
  {Deutsch}}]{norris2012enhanced}%
  \BibitemOpen
  \bibfield  {author} {\bibinfo {author} {\bibfnamefont {L.~M.}\ \bibnamefont
  {Norris}}, \bibinfo {author} {\bibfnamefont {C.~M.}\ \bibnamefont {Trail}},
  \bibinfo {author} {\bibfnamefont {P.~S.}\ \bibnamefont {Jessen}},\ and\
  \bibinfo {author} {\bibfnamefont {I.~H.}\ \bibnamefont {Deutsch}},\
  }\bibfield  {title} {\bibinfo {title} {Enhanced squeezing of a collective
  spin via control of its qudit subsystems},\ }\href@noop {} {\bibfield
  {journal} {\bibinfo  {journal} {Phys. Rev. Lett.}\ }\textbf {\bibinfo
  {volume} {109}},\ \bibinfo {pages} {173603} (\bibinfo {year}
  {2012})}\BibitemShut {NoStop}%
\bibitem [{\citenamefont {Kurucz}\ and\ \citenamefont
  {M{\o}lmer}(2010)}]{kurucz2010multilevel}%
  \BibitemOpen
  \bibfield  {author} {\bibinfo {author} {\bibfnamefont {Z.}~\bibnamefont
  {Kurucz}}\ and\ \bibinfo {author} {\bibfnamefont {K.}~\bibnamefont
  {M{\o}lmer}},\ }\bibfield  {title} {\bibinfo {title} {Multilevel
  holstein-primakoff approximation and its application to atomic spin squeezing
  and ensemble quantum memories},\ }\href@noop {} {\bibfield  {journal}
  {\bibinfo  {journal} {Phys. Rev. A}\ }\textbf {\bibinfo {volume} {81}},\
  \bibinfo {pages} {032314} (\bibinfo {year} {2010})}\BibitemShut {NoStop}%
\bibitem [{\citenamefont {Masson}\ \emph {et~al.}(2017)\citenamefont {Masson},
  \citenamefont {Barrett},\ and\ \citenamefont {Parkins}}]{masson2017cavity}%
  \BibitemOpen
  \bibfield  {author} {\bibinfo {author} {\bibfnamefont {S.~J.}\ \bibnamefont
  {Masson}}, \bibinfo {author} {\bibfnamefont {M.~D.}\ \bibnamefont
  {Barrett}},\ and\ \bibinfo {author} {\bibfnamefont {S.}~\bibnamefont
  {Parkins}},\ }\bibfield  {title} {\bibinfo {title} {Cavity qed engineering of
  spin dynamics and squeezing in a spinor gas},\ }\href@noop {} {\bibfield
  {journal} {\bibinfo  {journal} {Phys. Rev. Lett.}\ }\textbf {\bibinfo
  {volume} {119}},\ \bibinfo {pages} {213601} (\bibinfo {year}
  {2017})}\BibitemShut {NoStop}%
\bibitem [{\citenamefont {Sundar}\ \emph {et~al.}(2023)\citenamefont {Sundar},
  \citenamefont {Barberena}, \citenamefont {Pi\~neiro Orioli}, \citenamefont
  {Chu}, \citenamefont {Thompson}, \citenamefont {Rey},\ and\ \citenamefont
  {Lewis-Swan}}]{sundar2023bosonic}%
  \BibitemOpen
  \bibfield  {author} {\bibinfo {author} {\bibfnamefont {B.}~\bibnamefont
  {Sundar}}, \bibinfo {author} {\bibfnamefont {D.}~\bibnamefont {Barberena}},
  \bibinfo {author} {\bibfnamefont {A.}~\bibnamefont {Pi\~neiro Orioli}},
  \bibinfo {author} {\bibfnamefont {A.}~\bibnamefont {Chu}}, \bibinfo {author}
  {\bibfnamefont {J.~K.}\ \bibnamefont {Thompson}}, \bibinfo {author}
  {\bibfnamefont {A.~M.}\ \bibnamefont {Rey}},\ and\ \bibinfo {author}
  {\bibfnamefont {R.~J.}\ \bibnamefont {Lewis-Swan}},\ }\bibfield  {title}
  {\bibinfo {title} {Bosonic pair production and squeezing for optical phase
  measurements in long-lived dipoles coupled to a cavity},\ }\href@noop {}
  {\bibfield  {journal} {\bibinfo  {journal} {Phys. Rev. Lett}\ }\textbf
  {\bibinfo {volume} {130}},\ \bibinfo {pages} {113202} (\bibinfo {year}
  {2023})}\BibitemShut {NoStop}%
\bibitem [{\citenamefont {Gross}\ \emph {et~al.}(2011)\citenamefont {Gross},
  \citenamefont {Strobel}, \citenamefont {Nicklas}, \citenamefont {Zibold},
  \citenamefont {Bar-Gill}, \citenamefont {Kurizki},\ and\ \citenamefont
  {Oberthaler}}]{gross2011atomic}%
  \BibitemOpen
  \bibfield  {author} {\bibinfo {author} {\bibfnamefont {C.}~\bibnamefont
  {Gross}}, \bibinfo {author} {\bibfnamefont {H.}~\bibnamefont {Strobel}},
  \bibinfo {author} {\bibfnamefont {E.}~\bibnamefont {Nicklas}}, \bibinfo
  {author} {\bibfnamefont {T.}~\bibnamefont {Zibold}}, \bibinfo {author}
  {\bibfnamefont {N.}~\bibnamefont {Bar-Gill}}, \bibinfo {author}
  {\bibfnamefont {G.}~\bibnamefont {Kurizki}},\ and\ \bibinfo {author}
  {\bibfnamefont {M.~K.}\ \bibnamefont {Oberthaler}},\ }\bibfield  {title}
  {\bibinfo {title} {Atomic homodyne detection of continuous-variable entangled
  twin-atom states},\ }\href@noop {} {\bibfield  {journal} {\bibinfo  {journal}
  {Nature (London)}\ }\textbf {\bibinfo {volume} {480}},\ \bibinfo {pages}
  {219} (\bibinfo {year} {2011})}\BibitemShut {NoStop}%
\bibitem [{\citenamefont {L{\"u}cke}\ \emph {et~al.}(2011)\citenamefont
  {L{\"u}cke}, \citenamefont {Scherer}, \citenamefont {Kruse}, \citenamefont
  {Pezz{\'e}}, \citenamefont {Deuretzbacher}, \citenamefont {Hyllus},
  \citenamefont {Peise}, \citenamefont {Ertmer}, \citenamefont {Arlt},
  \citenamefont {Santos}, \citenamefont {Smerzi},\ and\ \citenamefont
  {Klempt}}]{lucke2011twin}%
  \BibitemOpen
  \bibfield  {author} {\bibinfo {author} {\bibfnamefont {B.}~\bibnamefont
  {L{\"u}cke}}, \bibinfo {author} {\bibfnamefont {M.}~\bibnamefont {Scherer}},
  \bibinfo {author} {\bibfnamefont {J.}~\bibnamefont {Kruse}}, \bibinfo
  {author} {\bibfnamefont {L.}~\bibnamefont {Pezz{\'e}}}, \bibinfo {author}
  {\bibfnamefont {F.}~\bibnamefont {Deuretzbacher}}, \bibinfo {author}
  {\bibfnamefont {P.}~\bibnamefont {Hyllus}}, \bibinfo {author} {\bibfnamefont
  {J.}~\bibnamefont {Peise}}, \bibinfo {author} {\bibfnamefont
  {W.}~\bibnamefont {Ertmer}}, \bibinfo {author} {\bibfnamefont
  {J.}~\bibnamefont {Arlt}}, \bibinfo {author} {\bibfnamefont {L.}~\bibnamefont
  {Santos}}, \bibinfo {author} {\bibfnamefont {A.}~\bibnamefont {Smerzi}},\
  and\ \bibinfo {author} {\bibfnamefont {C.}~\bibnamefont {Klempt}},\
  }\bibfield  {title} {\bibinfo {title} {Twin matter waves for interferometry
  beyond the classical limit},\ }\href@noop {} {\bibfield  {journal} {\bibinfo
  {journal} {Science}\ }\textbf {\bibinfo {volume} {334}},\ \bibinfo {pages}
  {773} (\bibinfo {year} {2011})}\BibitemShut {NoStop}%
\bibitem [{\citenamefont {Bookjans}\ \emph {et~al.}(2011)\citenamefont
  {Bookjans}, \citenamefont {Hamley},\ and\ \citenamefont
  {Chapman}}]{bookjans2011}%
  \BibitemOpen
  \bibfield  {author} {\bibinfo {author} {\bibfnamefont {E.~M.}\ \bibnamefont
  {Bookjans}}, \bibinfo {author} {\bibfnamefont {C.~D.}\ \bibnamefont
  {Hamley}},\ and\ \bibinfo {author} {\bibfnamefont {M.~S.}\ \bibnamefont
  {Chapman}},\ }\bibfield  {title} {\bibinfo {title} {Strong quantum spin
  correlations observed in atomic spin mixing},\ }\href
  {https://doi.org/10.1103/PhysRevLett.107.210406} {\bibfield  {journal}
  {\bibinfo  {journal} {Phys. Rev. Lett.}\ }\textbf {\bibinfo {volume} {107}},\
  \bibinfo {pages} {210406} (\bibinfo {year} {2011})}\BibitemShut {NoStop}%
\bibitem [{\citenamefont {Black}\ \emph {et~al.}(2007)\citenamefont {Black},
  \citenamefont {Gomez}, \citenamefont {Turner}, \citenamefont {Jung},\ and\
  \citenamefont {Lett}}]{Black2007}%
  \BibitemOpen
  \bibfield  {author} {\bibinfo {author} {\bibfnamefont {A.~T.}\ \bibnamefont
  {Black}}, \bibinfo {author} {\bibfnamefont {E.}~\bibnamefont {Gomez}},
  \bibinfo {author} {\bibfnamefont {L.~D.}\ \bibnamefont {Turner}}, \bibinfo
  {author} {\bibfnamefont {S.}~\bibnamefont {Jung}},\ and\ \bibinfo {author}
  {\bibfnamefont {P.~D.}\ \bibnamefont {Lett}},\ }\bibfield  {title} {\bibinfo
  {title} {Spinor dynamics in an antiferromagnetic spin-1 condensate},\ }\href
  {https://doi.org/10.1103/PhysRevLett.99.070403} {\bibfield  {journal}
  {\bibinfo  {journal} {Phys. Rev. Lett.}\ }\textbf {\bibinfo {volume} {99}},\
  \bibinfo {pages} {070403} (\bibinfo {year} {2007})}\BibitemShut {NoStop}%
\bibitem [{\citenamefont {Zhao}\ \emph {et~al.}(2014)\citenamefont {Zhao},
  \citenamefont {Jiang}, \citenamefont {Tang}, \citenamefont {Webb},\ and\
  \citenamefont {Liu}}]{zhao2014}%
  \BibitemOpen
  \bibfield  {author} {\bibinfo {author} {\bibfnamefont {L.}~\bibnamefont
  {Zhao}}, \bibinfo {author} {\bibfnamefont {J.}~\bibnamefont {Jiang}},
  \bibinfo {author} {\bibfnamefont {T.}~\bibnamefont {Tang}}, \bibinfo {author}
  {\bibfnamefont {M.}~\bibnamefont {Webb}},\ and\ \bibinfo {author}
  {\bibfnamefont {Y.}~\bibnamefont {Liu}},\ }\bibfield  {title} {\bibinfo
  {title} {Dynamics in spinor condensates tuned by a microwave dressing
  field},\ }\href {https://doi.org/10.1103/PhysRevA.89.023608} {\bibfield
  {journal} {\bibinfo  {journal} {Phys. Rev. A}\ }\textbf {\bibinfo {volume}
  {89}},\ \bibinfo {pages} {023608} (\bibinfo {year} {2014})}\BibitemShut
  {NoStop}%
\bibitem [{\citenamefont {Qu}\ \emph {et~al.}(2020)\citenamefont {Qu},
  \citenamefont {Evrard}, \citenamefont {Dalibard},\ and\ \citenamefont
  {Gerbier}}]{qu_2020}%
  \BibitemOpen
  \bibfield  {author} {\bibinfo {author} {\bibfnamefont {A.}~\bibnamefont
  {Qu}}, \bibinfo {author} {\bibfnamefont {B.}~\bibnamefont {Evrard}}, \bibinfo
  {author} {\bibfnamefont {J.}~\bibnamefont {Dalibard}},\ and\ \bibinfo
  {author} {\bibfnamefont {F.}~\bibnamefont {Gerbier}},\ }\bibfield  {title}
  {\bibinfo {title} {Probing spin correlations in a bose-einstein condensate
  near the single-atom level},\ }\href
  {https://doi.org/10.1103/PhysRevLett.125.033401} {\bibfield  {journal}
  {\bibinfo  {journal} {Phys. Rev. Lett.}\ }\textbf {\bibinfo {volume} {125}},\
  \bibinfo {pages} {033401} (\bibinfo {year} {2020})}\BibitemShut {NoStop}%
\bibitem [{\citenamefont {Kim}\ \emph {et~al.}(2021)\citenamefont {Kim},
  \citenamefont {Hur}, \citenamefont {Huh}, \citenamefont {Choi},\ and\
  \citenamefont {Choi}}]{Kim2021}%
  \BibitemOpen
  \bibfield  {author} {\bibinfo {author} {\bibfnamefont {K.}~\bibnamefont
  {Kim}}, \bibinfo {author} {\bibfnamefont {J.}~\bibnamefont {Hur}}, \bibinfo
  {author} {\bibfnamefont {S.~J.}\ \bibnamefont {Huh}}, \bibinfo {author}
  {\bibfnamefont {S.}~\bibnamefont {Choi}},\ and\ \bibinfo {author}
  {\bibfnamefont {J.-y.}\ \bibnamefont {Choi}},\ }\bibfield  {title} {\bibinfo
  {title} {Emission of spin-correlated matter-wave jets from spinor
  bose-einstein condensates},\ }\href
  {https://doi.org/10.1103/PhysRevLett.127.043401} {\bibfield  {journal}
  {\bibinfo  {journal} {Phys. Rev. Lett.}\ }\textbf {\bibinfo {volume} {127}},\
  \bibinfo {pages} {043401} (\bibinfo {year} {2021})}\BibitemShut {NoStop}%
\bibitem [{\citenamefont {Julsgaard}\ \emph {et~al.}(2001)\citenamefont
  {Julsgaard}, \citenamefont {Kozhekin},\ and\ \citenamefont
  {Polzik}}]{julsgaard2001experimental}%
  \BibitemOpen
  \bibfield  {author} {\bibinfo {author} {\bibfnamefont {B.}~\bibnamefont
  {Julsgaard}}, \bibinfo {author} {\bibfnamefont {A.}~\bibnamefont
  {Kozhekin}},\ and\ \bibinfo {author} {\bibfnamefont {E.~S.}\ \bibnamefont
  {Polzik}},\ }\bibfield  {title} {\bibinfo {title} {Experimental long-lived
  entanglement of two macroscopic objects},\ }\href@noop {} {\bibfield
  {journal} {\bibinfo  {journal} {Nature (London)}\ }\textbf {\bibinfo {volume}
  {413}},\ \bibinfo {pages} {400} (\bibinfo {year} {2001})}\BibitemShut
  {NoStop}%
\bibitem [{\citenamefont {Polzik}\ and\ \citenamefont
  {Ye}(2016)}]{polzik2016entanglement}%
  \BibitemOpen
  \bibfield  {author} {\bibinfo {author} {\bibfnamefont {E.~S.}\ \bibnamefont
  {Polzik}}\ and\ \bibinfo {author} {\bibfnamefont {J.}~\bibnamefont {Ye}},\
  }\bibfield  {title} {\bibinfo {title} {Entanglement and spin squeezing in a
  network of distant optical lattice clocks},\ }\href@noop {} {\bibfield
  {journal} {\bibinfo  {journal} {Phys. Rev. A}\ }\textbf {\bibinfo {volume}
  {93}},\ \bibinfo {pages} {021404(R)} (\bibinfo {year} {2016})}\BibitemShut
  {NoStop}%
\bibitem [{\citenamefont {Ludlow}\ \emph {et~al.}(2015)\citenamefont {Ludlow},
  \citenamefont {Boyd}, \citenamefont {Ye}, \citenamefont {Peik},\ and\
  \citenamefont {Schmidt}}]{ludlow2015optical}%
  \BibitemOpen
  \bibfield  {author} {\bibinfo {author} {\bibfnamefont {A.~D.}\ \bibnamefont
  {Ludlow}}, \bibinfo {author} {\bibfnamefont {M.~M.}\ \bibnamefont {Boyd}},
  \bibinfo {author} {\bibfnamefont {J.}~\bibnamefont {Ye}}, \bibinfo {author}
  {\bibfnamefont {E.}~\bibnamefont {Peik}},\ and\ \bibinfo {author}
  {\bibfnamefont {P.~O.}\ \bibnamefont {Schmidt}},\ }\bibfield  {title}
  {\bibinfo {title} {Optical atomic clocks},\ }\href@noop {} {\bibfield
  {journal} {\bibinfo  {journal} {Rev. Mod. Phys.}\ }\textbf {\bibinfo {volume}
  {87}},\ \bibinfo {pages} {637} (\bibinfo {year} {2015})}\BibitemShut
  {NoStop}%
\bibitem [{SM()}]{SM}%
  \BibitemOpen
  \href@noop {} {\bibinfo {title} {{See Supplemental Material at [URL will be
  inserted by publisher] which includes Ref.~\cite{vitagliano2011spin} for
  derivations of master equation, mean-field physics, squeezing, transfering
  the squeezing, and numerical benchmarks}}}\BibitemShut {NoStop}%
\bibitem [{\citenamefont {Vitagliano}\ \emph {et~al.}(2011)\citenamefont
  {Vitagliano}, \citenamefont {Hyllus}, \citenamefont {Egusquiza},\ and\
  \citenamefont {T{\'o}th}}]{vitagliano2011spin}%
  \BibitemOpen
  \bibfield  {author} {\bibinfo {author} {\bibfnamefont {G.}~\bibnamefont
  {Vitagliano}}, \bibinfo {author} {\bibfnamefont {P.}~\bibnamefont {Hyllus}},
  \bibinfo {author} {\bibfnamefont {I.~L.}\ \bibnamefont {Egusquiza}},\ and\
  \bibinfo {author} {\bibfnamefont {G.}~\bibnamefont {T{\'o}th}},\ }\bibfield
  {title} {\bibinfo {title} {Spin squeezing inequalities for arbitrary spin},\
  }\href@noop {} {\bibfield  {journal} {\bibinfo  {journal} {Phys. Rev. Lett.}\
  }\textbf {\bibinfo {volume} {107}},\ \bibinfo {pages} {240502} (\bibinfo
  {year} {2011})}\BibitemShut {NoStop}%
\bibitem [{\citenamefont {Vitagliano}\ \emph {et~al.}(2014)\citenamefont
  {Vitagliano}, \citenamefont {Apellaniz}, \citenamefont {Egusquiza},\ and\
  \citenamefont {T{\'o}th}}]{vitagliano2014spin}%
  \BibitemOpen
  \bibfield  {author} {\bibinfo {author} {\bibfnamefont {G.}~\bibnamefont
  {Vitagliano}}, \bibinfo {author} {\bibfnamefont {I.}~\bibnamefont
  {Apellaniz}}, \bibinfo {author} {\bibfnamefont {I.~L.}\ \bibnamefont
  {Egusquiza}},\ and\ \bibinfo {author} {\bibfnamefont {G.}~\bibnamefont
  {T{\'o}th}},\ }\bibfield  {title} {\bibinfo {title} {Spin squeezing and
  entanglement for an arbitrary spin},\ }\href@noop {} {\bibfield  {journal}
  {\bibinfo  {journal} {Phys. Rev. A}\ }\textbf {\bibinfo {volume} {89}},\
  \bibinfo {pages} {032307} (\bibinfo {year} {2014})}\BibitemShut {NoStop}%
\bibitem [{\citenamefont {Lieu}\ \emph {et~al.}(2020)\citenamefont {Lieu},
  \citenamefont {Belyansky}, \citenamefont {Young}, \citenamefont {Lundgren},
  \citenamefont {Albert},\ and\ \citenamefont {Gorshkov}}]{lieu2020symmetry}%
  \BibitemOpen
  \bibfield  {author} {\bibinfo {author} {\bibfnamefont {S.}~\bibnamefont
  {Lieu}}, \bibinfo {author} {\bibfnamefont {R.}~\bibnamefont {Belyansky}},
  \bibinfo {author} {\bibfnamefont {J.~T.}\ \bibnamefont {Young}}, \bibinfo
  {author} {\bibfnamefont {R.}~\bibnamefont {Lundgren}}, \bibinfo {author}
  {\bibfnamefont {V.~V.}\ \bibnamefont {Albert}},\ and\ \bibinfo {author}
  {\bibfnamefont {A.~V.}\ \bibnamefont {Gorshkov}},\ }\bibfield  {title}
  {\bibinfo {title} {Symmetry breaking and error correction in open quantum
  systems},\ }\href@noop {} {\bibfield  {journal} {\bibinfo  {journal} {Phys.
  Rev. Lett.}\ }\textbf {\bibinfo {volume} {125}},\ \bibinfo {pages} {240405}
  (\bibinfo {year} {2020})}\BibitemShut {NoStop}%
\bibitem [{\citenamefont {Prosen}(2008)}]{prosen2008third}%
  \BibitemOpen
  \bibfield  {author} {\bibinfo {author} {\bibfnamefont {T.}~\bibnamefont
  {Prosen}},\ }\bibfield  {title} {\bibinfo {title} {Third quantization: a
  general method to solve master equations for quadratic open fermi systems},\
  }\href@noop {} {\bibfield  {journal} {\bibinfo  {journal} {New J. Phys.}\
  }\textbf {\bibinfo {volume} {10}},\ \bibinfo {pages} {043026} (\bibinfo
  {year} {2008})}\BibitemShut {NoStop}%
\bibitem [{\citenamefont {Bu{\v{c}}a}\ and\ \citenamefont
  {Prosen}(2012)}]{buvca2012note}%
  \BibitemOpen
  \bibfield  {author} {\bibinfo {author} {\bibfnamefont {B.}~\bibnamefont
  {Bu{\v{c}}a}}\ and\ \bibinfo {author} {\bibfnamefont {T.}~\bibnamefont
  {Prosen}},\ }\bibfield  {title} {\bibinfo {title} {A note on symmetry
  reductions of the lindblad equation: transport in constrained open spin
  chains},\ }\href@noop {} {\bibfield  {journal} {\bibinfo  {journal} {New J.
  Phys.}\ }\textbf {\bibinfo {volume} {14}},\ \bibinfo {pages} {073007}
  (\bibinfo {year} {2012})}\BibitemShut {NoStop}%
\bibitem [{\citenamefont {Barthel}\ and\ \citenamefont
  {Zhang}(2022)}]{barthel2022solving}%
  \BibitemOpen
  \bibfield  {author} {\bibinfo {author} {\bibfnamefont {T.}~\bibnamefont
  {Barthel}}\ and\ \bibinfo {author} {\bibfnamefont {Y.}~\bibnamefont
  {Zhang}},\ }\bibfield  {title} {\bibinfo {title} {Solving quasi-free and
  quadratic lindblad master equations for open fermionic and bosonic systems},\
  }\href@noop {} {\bibfield  {journal} {\bibinfo  {journal} {J. Stat. Mech.}\
  ,\ \bibinfo {pages} {113101}} (\bibinfo {year} {2022})}\BibitemShut {NoStop}%
\bibitem [{\citenamefont {Omanakuttan}\ \emph {et~al.}(2021)\citenamefont
  {Omanakuttan}, \citenamefont {Mitra}, \citenamefont {Martin},\ and\
  \citenamefont {Deutsch}}]{omanakuttan2021quantum}%
  \BibitemOpen
  \bibfield  {author} {\bibinfo {author} {\bibfnamefont {S.}~\bibnamefont
  {Omanakuttan}}, \bibinfo {author} {\bibfnamefont {A.}~\bibnamefont {Mitra}},
  \bibinfo {author} {\bibfnamefont {M.~J.}\ \bibnamefont {Martin}},\ and\
  \bibinfo {author} {\bibfnamefont {I.~H.}\ \bibnamefont {Deutsch}},\
  }\bibfield  {title} {\bibinfo {title} {Quantum optimal control of ten-level
  nuclear spin qudits in sr 87},\ }\href@noop {} {\bibfield  {journal}
  {\bibinfo  {journal} {Phys. Rev. A}\ }\textbf {\bibinfo {volume} {104}},\
  \bibinfo {pages} {L060401} (\bibinfo {year} {2021})}\BibitemShut {NoStop}%
\bibitem [{\citenamefont {Pezz{\`e}}\ \emph {et~al.}(2018)\citenamefont
  {Pezz{\`e}}, \citenamefont {Smerzi}, \citenamefont {Oberthaler},
  \citenamefont {Schmied},\ and\ \citenamefont {Treutlein}}]{pezze2018quantum}%
  \BibitemOpen
  \bibfield  {author} {\bibinfo {author} {\bibfnamefont {L.}~\bibnamefont
  {Pezz{\`e}}}, \bibinfo {author} {\bibfnamefont {A.}~\bibnamefont {Smerzi}},
  \bibinfo {author} {\bibfnamefont {M.~K.}\ \bibnamefont {Oberthaler}},
  \bibinfo {author} {\bibfnamefont {R.}~\bibnamefont {Schmied}},\ and\ \bibinfo
  {author} {\bibfnamefont {P.}~\bibnamefont {Treutlein}},\ }\bibfield  {title}
  {\bibinfo {title} {Quantum metrology with nonclassical states of atomic
  ensembles},\ }\href@noop {} {\bibfield  {journal} {\bibinfo  {journal} {Rev.
  Mod. Phys.}\ }\textbf {\bibinfo {volume} {90}},\ \bibinfo {pages} {035005}
  (\bibinfo {year} {2018})}\BibitemShut {NoStop}%
\end{thebibliography}%


\begin{thebibliography}{2}%
\makeatletter
\providecommand \@ifxundefined [1]{%
 \@ifx{#1\undefined}
}%
\providecommand \@ifnum [1]{%
 \ifnum #1\expandafter \@firstoftwo
 \else \expandafter \@secondoftwo
 \fi
}%
\providecommand \@ifx [1]{%
 \ifx #1\expandafter \@firstoftwo
 \else \expandafter \@secondoftwo
 \fi
}%
\providecommand \natexlab [1]{#1}%
\providecommand \enquote  [1]{``#1''}%
\providecommand \bibnamefont  [1]{#1}%
\providecommand \bibfnamefont [1]{#1}%
\providecommand \citenamefont [1]{#1}%
\providecommand \href@noop [0]{\@secondoftwo}%
\providecommand \href [0]{\begingroup \@sanitize@url \@href}%
\providecommand \@href[1]{\@@startlink{#1}\@@href}%
\providecommand \@@href[1]{\endgroup#1\@@endlink}%
\providecommand \@sanitize@url [0]{\catcode `\\12\catcode `\$12\catcode
  `\&12\catcode `\#12\catcode `\^12\catcode `\_12\catcode `\%12\relax}%
\providecommand \@@startlink[1]{}%
\providecommand \@@endlink[0]{}%
\providecommand \url  [0]{\begingroup\@sanitize@url \@url }%
\providecommand \@url [1]{\endgroup\@href {#1}{\urlprefix }}%
\providecommand \urlprefix  [0]{URL }%
\providecommand \Eprint [0]{\href }%
\providecommand \doibase [0]{https://doi.org/}%
\providecommand \selectlanguage [0]{\@gobble}%
\providecommand \bibinfo  [0]{\@secondoftwo}%
\providecommand \bibfield  [0]{\@secondoftwo}%
\providecommand \translation [1]{[#1]}%
\providecommand \BibitemOpen [0]{}%
\providecommand \bibitemStop [0]{}%
\providecommand \bibitemNoStop [0]{.\EOS\space}%
\providecommand \EOS [0]{\spacefactor3000\relax}%
\providecommand \BibitemShut  [1]{\csname bibitem#1\endcsname}%
\let\auto@bib@innerbib\@empty
%</preamble>
\bibitem [{\citenamefont {Vitagliano}\ \emph {et~al.}(2011)\citenamefont
  {Vitagliano}, \citenamefont {Hyllus}, \citenamefont {Egusquiza},\ and\
  \citenamefont {T{\'o}th}}]{vitagliano2011spin}%
  \BibitemOpen
  \bibfield  {author} {\bibinfo {author} {\bibfnamefont {G.}~\bibnamefont
  {Vitagliano}}, \bibinfo {author} {\bibfnamefont {P.}~\bibnamefont {Hyllus}},
  \bibinfo {author} {\bibfnamefont {I.~L.}\ \bibnamefont {Egusquiza}},\ and\
  \bibinfo {author} {\bibfnamefont {G.}~\bibnamefont {T{\'o}th}},\ }\bibfield
  {title} {\bibinfo {title} {Spin squeezing inequalities for arbitrary spin},\
  }\href@noop {} {\bibfield  {journal} {\bibinfo  {journal} {Phys. Rev. Lett.}\
  }\textbf {\bibinfo {volume} {107}},\ \bibinfo {pages} {240502} (\bibinfo
  {year} {2011})}\BibitemShut {NoStop}%
\bibitem [{\citenamefont {Sundar}\ \emph {et~al.}()\citenamefont {Sundar},
  \citenamefont {Barberena}, \citenamefont {Rey},\ and\ \citenamefont
  {Pi\~neiro Orioli}}]{pra}%
  \BibitemOpen
  \bibfield  {author} {\bibinfo {author} {\bibfnamefont {B.}~\bibnamefont
  {Sundar}}, \bibinfo {author} {\bibfnamefont {D.}~\bibnamefont {Barberena}},
  \bibinfo {author} {\bibfnamefont {A.~M.}\ \bibnamefont {Rey}},\ and\ \bibinfo
  {author} {\bibfnamefont {A.}~\bibnamefont {Pi\~neiro Orioli}},\ }\bibfield
  {title} {\bibinfo {title} {Driven-dissipative four-mode squeezing of
  multilevel atoms in an optical cavity},\ }\href@noop {} {\bibinfo  {journal}
  {Article in preparation}\ }\BibitemShut {NoStop}%
\end{thebibliography}%
\end{document}

\title{Supplementary Material for: Squeezing multilevel atoms in a dark state via cavity superradiance}

\author{Bhuvanesh Sundar$^*$}
\affiliation{JILA, NIST, Department of Physics, University of Colorado, Boulder, CO 80309, USA}
\affiliation{Center for Theory of Quantum Matter, University of Colorado, Boulder, CO 80309, USA}
\thanks{Now at Rigetti Computing, Berkeley, CA 94710, USA}

\author{Diego Barberena}
\affiliation{JILA, NIST, Department of Physics, University of Colorado, Boulder, CO 80309, USA}
\affiliation{Center for Theory of Quantum Matter, University of Colorado, Boulder, CO 80309, USA}

\author{Ana Maria Rey}
\affiliation{JILA, NIST, Department of Physics, University of Colorado, Boulder, CO 80309, USA}
\affiliation{Center for Theory of Quantum Matter, University of Colorado, Boulder, CO 80309, USA}

\author{Asier Pi\~neiro Orioli}
\affiliation{JILA, NIST, Department of Physics, University of Colorado, Boulder, CO 80309, USA}
\affiliation{Center for Theory of Quantum Matter, University of Colorado, Boulder, CO 80309, USA}
\maketitle

\section{Deriving the effective multilevel spin model}\label{app: effective master equation}
In the main text, we considered coherently driven atoms in a cavity undergoing superradiant emission [Eq.(1)] in the bad-cavity limit. Here, we present a detailed discussion of the full atom-light master equation describing this system and show that it can be reduced to the effective spin master equation considered in the main text.

The dynamics of the atom-light system is modeled by the Lindblad master equation,
\begin{equation}\label{eqn: master}
\hbar\frac{d\rho}{dt} = -i[\hat H_{\rm tot}, \rho] + \mathcal{L}_{\rm cav}[\rho].
\end{equation}
Here, $\hat H_{\rm tot} = \hat H_A + \hat H_L + \hat H_{AL}$ is a Hamiltonian including contributions from the atoms, cavity modes, atom-light coupling, and external driving:
\begin{align}\label{eqn: H spin-photon}
& \hat H_A = \hbar\omega \hat n_e,\nonumber\\
& \hat H_L = \sum_{\alpha} \hbar\omega \ha_\alpha\+ \ha_\alpha^{\phantom\dagger} + \frac{i\hbar\epsilon_\alpha}{2}(\ha_\alpha\+ e^{i\omega t} - \ha_\alpha^{\phantom\dagger} e^{-i\omega t}),\nonumber\\
& \hat H_{AL} = \hbar g \sum_\alpha \ha_\alpha \hat{D}_\alpha^+ + {\rm h.c.} ,
\end{align}
and $\mathcal{L}_{\rm cav}[\rho]$ describes the dissipation given by the jump operators $\hat L_\alpha = \sqrt{\kappa}\hat a_\alpha$ due to leakage of photons out of the cavity at rate $\kappa$:
\begin{equation}\label{eqn: Lcav}
\mathcal{L}_{\rm cav}[\rho] = \hbar\kappa\sum_{\alpha} \left(\ha_\alpha^{\phantom\dagger} \rho \ha_\alpha\+ - \frac{1}{2}\ha_\alpha\+\ha_\alpha^{\phantom\dagger} \rho - \frac{1}{2}\rho\ha_\alpha\+\ha_\alpha^{\phantom\dagger}\right) .
\end{equation}
In the above equations, $\hat n_e$ is the occupation in the excited manifold, and $\ha_\alpha$ annihilates a photon in the cavity mode with polarization $\alpha$.

It is convenient to move to a rotating frame that rotates at the frequency $\omega$. In this frame, the atomic angular frequency, cavity frequency, and laser frequency are shifted by $\omega$, yielding the Hamiltonian
\begin{equation}
\hat H_{\rm tot} = \sum_\alpha \frac{i\hbar\epsilon_\alpha}{2}(\ha_\alpha\+ - \ha_\alpha^{\phantom\dagger}) + \hbar g \left( \ha_\alpha \hat{D}_\alpha^+ + {\rm h.c.} \right),
\end{equation}
and the same Lindblad jump operators as before.

The master equation for the photon operators is
\begin{equation}
\partial_t \braket{\ha_\alpha} = \braket{ -\frac{\kappa}{2} \ha_\alpha - i g \hat D_\alpha^- + \frac{\epsilon_\alpha}{2} }.
\end{equation}
Assuming the bad cavity limit, $\kappa \gg g\sqrt{N}$, the photons' evolution follows the spins,
\begin{equation}\label{eqn: a in terms of spins}
\ha_\alpha \rightarrow \frac{i\epsilon_\alpha + 2g \hat D_\alpha^-}{i\kappa}.
\end{equation}
We adiabatically eliminate the photons by substituting Eq.~\eqref{eqn: a in terms of spins} into Eqs.~\eqref{eqn: H spin-photon} and~\eqref{eqn: Lcav}, and obtain
\begin{equation}\label{eqn: Heff and Leff}
\hat H_{\rm eff} = 0,\quad \hat L_{\rm eff,\alpha} = \frac{\sqrt{\hbar\kappa}(i\epsilon_\alpha + 2g \hat D_\alpha^-)}{i\kappa}.
\end{equation}
Defining $\Gamma = \frac{4g^2}{\kappa}$ and $\Omega_\alpha = \frac{2\epsilon_\alpha g}{\kappa}$, we find that $\hat L_{\rm eff,\alpha} = -i\sqrt{\hbar\Gamma}\left(\hat D^-_\alpha + i\frac{\Omega_\alpha}{\Gamma}\right) = -i\sqrt{\hbar\Gamma} \hat{\mathscr{D}}^-_\alpha$. The factor $-i$ in front is irrelevant, therefore we drop it hereafter.

The master equation due to $\hat{\mathscr{D}}^-_\alpha$,
\begin{equation}
\dot{\rho} = \hbar\Gamma \sum_\alpha \hat{\mathscr{D}}^-_\alpha \rho \hat{\mathscr{D}}^+_\alpha - \frac{1}{2}\{ \hat{\mathscr{D}}^+_\alpha \hat{\mathscr{D}}^-_\alpha, \rho\},
\end{equation}
is equivalent to the master equation with $\hat H$ and $\mathcal{L}[\rho]$ given in Eq.(1) of the main text.

\section{Mean-field evolution}
We described the mean-field evolution of the Bloch vector in terms of a superradiance potential in the main text. Here, we derive the mean-field equations, and obtain the simple description in terms of $V_\beta(\theta)$.

The master equation for $\braket{\hat S^\mu_{m,+1}}$, $\mu \in \{x,y,z\}$, is
\begin{align}\label{eqn: MF bloch vector}
\frac{d}{dt}\braket{\hat S^\mu_{m,+1}} = & -i\Omega \braket{ [\hat S^\mu_{m,+1}, \hat D^x_{+1}] } \nonumber\\
& + \frac{\Gamma}{2} \braket{ \hat D^+_{+1} [\hat S^\mu_{m,+1}, \hat D^-_{+1}] } + \braket{ [\hat D^+_{+1} , \hat S^\mu_{m,+1}] \hat D^-_{+1} }.
\end{align}
Making the mean-field approximation, $\braket{\ha \hb} \approx \braket{\ha}\braket{\hb}$, Eq.~\eqref{eqn: MF bloch vector} simplifies to
\begin{align}\label{eqn: MF bloch vector 2}
\frac{d}{dt}\braket{\hat S^\mu_{m,+1}} = & -i(\Omega - \Gamma \braket{\hat D^y_{+1}})\braket{ [\hat S^\mu_{m,+1}, \hat D^x_{+1}] } \nonumber\\
& - i \Gamma \braket{ \hat D^x_{+1}} \braket{ [\hat S^\mu_{m,+1}, \hat D^y_{+1}] }.
\end{align}
Finally, realizing that the equation for the commutators, $[\hat S^\mu_{m,+1}, \hat D^\nu_{+1}] = i \sum_\lambda \epsilon_{\mu\nu\lambda} C_m^{+1} \hat S^\lambda_{m,+1}$, have the same form as the equation for a cross product, we can write Eq.~\eqref{eqn: MF bloch vector 2} as
\begin{equation}\label{eqn: cross product}
\dot{\vec{S}}_m = C_m^{+1}(\Omega - \Gamma \braket{\hat D^y_{+1}}, \Gamma\braket{\hat D^x_{+1}}, 0) \times \vec{S}_m.
\end{equation}

For the given initial condition, a parametric equation that solves Eq.~\eqref{eqn: cross product} is $\dot{\theta} = \Omega - \Gamma \braket{\hat D^y_{+1}}$ 
with $\vec{S}_m = |\vec{S}_m| (0, \sin[C_m^{+1} \theta], -\cos[C_m^{+1} \theta])$. We define a potential $V_\beta(\theta)$ such that
\begin{align}
\frac{\partial V_\beta(\theta)}{\partial\theta} &= \frac{1}{N}\braket{\Psi(\theta;\beta) \vert \hat D^y_{+1} \vert \Psi(\theta;\beta)} 
\nonumber\\
&= \frac{1}{N}\braket{G_\beta \vert e^{i\theta \hat D^x_{+1}} \hat D^y_{+1} e^{-i\theta \hat D^x_{+1}} \vert G_\beta} .
\end{align}
Using the commutation $[\sum_m \hat S^z_m, \hat D^x_{+1}] = i\hat D^y_{+1}$, it follows that
\begin{align}
V_\beta(\theta) &= c + \frac{1}{N}\sum_m \braket{G_\beta \vert e^{i\theta \hat D^x_{+1}} \hat S^z_m e^{-i\theta \hat D^x_{+1}} \vert G_\beta} \nonumber\\ &= c + \frac{1}{N}\sum_m \braket{\Psi(\theta;\beta) \vert \hat S^z_m \vert \Psi(\theta;\beta)}
\end{align}
where $c$ is an irrelevant additive constant.

\section{Equivalent bosonic description}
In the main text, we defined the Schwinger bosons $\hc_i$ as a unitary transformation of $\ha_i$, $\hc_i = \sum_j (M_{\theta_0})_{ij} \ha_j$. Here, we (a) explicitly define $\hc_i$, and (b) simultaneously give a general recipe to find a definition such that the jump operator (for any polarization and multilevel structure) has the form $\hat{\mathscr{D}}^- \approx \sqrt{N} x \hat X^c_1 + iy(\cos\phi \hat Y^c_1 + \sin\phi \hat Y^c_2)$ that we used in the main text [see Eq.~(3)].

\begin{enumerate}
\item 
By convention, we choose $\hc_0$ such that it annihilates particles in the ``condensate''. For the present case, this gives
\begin{align}
 \hc_0 = & \cos\frac{\beta}{2} \left(\cos\frac{\theta_0}{2\sqrt{3}} \ha_{g,-1/2} + i \sin\frac{\theta_0}{2\sqrt{3}} \ha_{e,1/2}\right) \nonumber\\ & + \sin\frac{\beta}{2}\left( \cos\frac{\theta_0}{2}\ha_{g,1/2} + i \sin\frac{\theta_0}{2} \ha_{e,3/2} \right).
 \end{align}
\item The next Schwinger boson is $\hc_1 = \sum_j M_{1j} \ha_j$. We choose $\hc_1$ via
\begin{equation}\label{eqn: Dx}
\hat D^x_{+1} = \sqrt{N}x \hat X^c_1
\end{equation}
in the HP approximation. This uniquely determines the coefficients $M_{1j}$ and thus $\hc_1$ as
\begin{align}
 \hc_1 = & \frac{\cos\frac{\beta}{2}}{\sqrt{\cos^2\frac{\beta}{2}+3\sin^2\frac{\beta}{2}}} \left( i\sin\frac{\theta}{2\sqrt{3}} \ha_{g,-1/2} + \cos\frac{\theta}{2\sqrt{3}} \ha_{e,1/2}\right)
\nonumber\\ 
& + \frac{\sqrt{3}\sin\frac{\beta}{2}}{\sqrt{\cos^2\frac{\beta}{2}+3\sin^2\frac{\beta}{2}}} \left( i \sin\frac{\theta}{2} \ha_{g,1/2} + \cos\frac{\theta}{2} \ha_{e,3/2} 
 \right).
\end{align}
Moreover, by equating the variance of $\hat D^x_{+1}$ in the mean-field state with the variance of $\sqrt{N}x \hat X^c_1$ in the coherent vacuum of $\hc_1$, we find that
\begin{equation}
x = \sqrt{\frac{2}{N}} (\Delta \hat D^x_{+1})_{MF}.
\end{equation}
For the present case, this value is $x = \sqrt{ \frac{\cos^2(\beta/2)}{6} + \frac{\sin^2(\beta/2)}{2} }$.
\item Next, we define $\hc_2$ via $\hat{\mathscr{D}}^y_{+1}$.
In the HP approximation, $\hat{\mathscr{D}}^y_{+1}$ reduces to
\begin{equation}\label{eqn: Dy}
\hat{\mathscr{D}}^y_{+1} = \sqrt{N}(\alpha \hat X^c_1 + \alpha' Y^c_1 + {\rm remaining}).
\end{equation}
where the relations
\begin{align}\label{eqn: alpha'}
\alpha = &\frac{1}{N x}\braket{ \{ \hat{D}^x_{+1}, \hat{\mathscr{D}}^y_{+1} \} }_{MF}, \nonumber\\
\alpha' = &\frac{-i}{N x}\braket{ [ \hat{D}^x_{+1}, \hat{\mathscr{D}}^y_{+1} ] }_{MF}
\end{align}
can be shown by direct substitution into Eqs.~\eqref{eqn: Dx} and~\eqref{eqn: Dy}. 
We define $\hc_2$ such that the `remaining' terms in $\hat{\mathscr{D}}^y_{+1}$ are proportional to $\hat Y^c_2$. To determine $\alpha$, note the identity $\braket{\{ \hat D^x_{+1}, \hat{\mathscr{D}}^y_{+1} \}}_{MF} = 0$, therefore $\alpha = 0$. Therefore, $\hat{\mathscr{D}}^y_{+1}$ in the HP approximation is $\hat{\mathscr{D}}^y_{+1} = \sqrt{N}(\alpha' \hat Y^c_1 + \alpha'' \hat Y^c_2)$, and we defined
\begin{align}
\alpha' \equiv y\cos\phi \nonumber\\
\alpha'' \equiv y\sin\phi
\end{align}
in the main text. Similar to $x$, we have that $y = \sqrt{\frac{2}{N}} (\Delta \hat D^y_{+1})_{MF}$. Using this fact and the expression for $\alpha'\equiv y\cos\phi$ in Eq.~\eqref{eqn: alpha'}, we obtain
\begin{align}
y &= \sqrt{\frac{\cos^2\frac{\beta}{2}}{6} + \frac{\sin^2\frac{\beta}{2}}{2} - \frac{1}{2}\left(\frac{\cos^2\frac{\beta}{2} \sin\frac{\theta}{\sqrt{3}}}{\sqrt{3}} + \sin^2\frac{\beta}{2}\sin\theta\right)^2 } \nonumber\\ &= \sqrt{x^2 - \frac{2\Omega^2}{(N\Gamma)^2}}, \\
\cos\phi &= \frac{1}{6xy}\left(\cos^2\frac{\beta}{2}\cos\frac{\theta}{\sqrt{3}} + 3\sin^2\frac{\beta}{2}\cos\theta \right),
\end{align}
where $\Omega$ is the coherent driving strength.

Finally, from the HP expansion $\hat{\mathscr{D}}^y_{+1} = y(\cos\phi \hat Y^c_1 + \sin\phi \hat Y^c_2)$, we obtain the definition of $\hc_2$,
\begin{widetext}
\begin{align}
\label{eqn: remaining}
\hc_2 = &\frac{\sin\beta}{2\sqrt{2} y\sin\phi}\left(\sin\theta - \frac{1}{\sqrt{3}}\sin\frac{\theta}{\sqrt{3}}\right) \nonumber\\
& \times \left( i\sin\frac{\beta}{2}\cos\frac{\theta}{2\sqrt{3}}\ha_{g,-1/2} - i\cos\frac{\beta}{2}\cos\frac{\theta}{2}\ha_{g,1/2} - \sin\frac{\beta}{2}\sin\frac{\theta}{2\sqrt{3}}\ha_{e,1/2} + \cos\frac{\beta}{2}\sin\frac{\theta}{2}\ha_{e,3/2} \right) \nonumber\\
& + \frac{\sin\beta}{2\sqrt{2} y\sin\phi} \frac{ \cos\theta - \cos\frac{\theta}{\sqrt{3}} }{ \cos^2\frac{\beta}{2}+3\sin^2\frac{\beta}{2} } \nonumber\\
& \times \left( -i\sqrt{3}\sin\frac{\beta}{2}\sin\frac{\theta}{2\sqrt{3}}\ha_{g,-1/2} + i\cos\frac{\beta}{2}\sin\frac{\theta}{2}\ha_{g,1/2} - \sqrt{3}\sin\frac{\beta}{2}\cos\frac{\theta}{2\sqrt{3}}\ha_{e,1/2} + \cos\frac{\beta}{2}\cos\frac{\theta}{2}\ha_{e,3/2} \right) \\
\end{align}\end{widetext}
\item
The third Schwinger boson, $\hc_3$, is uniquely determined as the one that commutes with $\hc_0\+$, $\hc_1\+$, and $\hc_2\+$:
\begin{widetext}
\begin{align}
\hc_3 = &\frac{\sin\beta}{4\sqrt{3} xy\sin\phi}\left(\cos\theta - \cos\frac{\theta}{\sqrt{3}}\right) \nonumber\\
& \times \left( i\sin\frac{\beta}{2}\cos\frac{\theta}{2\sqrt{3}}\ha_{g,-1/2} - i\cos\frac{\beta}{2}\cos\frac{\theta}{2}\ha_{g,1/2} - \sin\frac{\beta}{2}\sin\frac{\theta}{2\sqrt{3}}\ha_{e,1/2} + \cos\frac{\beta}{2}\sin\frac{\theta}{2}\ha_{e,3/2} \right) \nonumber\\
& - \frac{\sin\beta}{4\sqrt{3} xy\sin\phi} \left( \sin\theta - \frac{1}{\sqrt{3}}\sin\frac{\theta}{\sqrt{3}} \right) \nonumber\\
& \times \left( -i\sqrt{3}\sin\frac{\beta}{2}\sin\frac{\theta}{2\sqrt{3}}\ha_{g,-1/2} + i\cos\frac{\beta}{2}\sin\frac{\theta}{2}\ha_{g,1/2} - \sqrt{3}\sin\frac{\beta}{2}\cos\frac{\theta}{2\sqrt{3}}\ha_{e,1/2} + \cos\frac{\beta}{2}\cos\frac{\theta}{2}\ha_{e,3/2} \right)
\end{align}
\end{widetext}
\end{enumerate}

\Rev{\section{Quantum noise in the HP approximation}}
The covariance matrix $\Sigma$ calculates the connected correlations between $\hat X^c_i$ and $\hat Y^c_i$. 
For the multilevel system considered, which has 3 HP bosons, it is defined as $\Sigma_{ij} = \langle \{ \hat{X}^c_i , \hat{X}^c_j \} \rangle_C$, $\Sigma_{i+3,j+3} = \langle \{ \hat{Y}^c_i , \hat{Y}^c_j \} \rangle_C$, and $\Sigma_{i,j+3} = \langle \{ \hat{X}^c_i , \hat{Y}^c_j \} \rangle_C$ for $i,j\in\{1,2,3\}$.
Note that $\Sigma$ is symmetric and by definition of the HP bosons we have $\langle \hat{X}^c_i \rangle = \langle \hat{Y}^c_i \rangle =0$, so we will drop the subscript `C' in the following.

The initial uncorrelated state is the vacuum of $\hc_{i>0}$, therefore the initial value of the covariance matrix is 
$\Sigma = \left( \begin{array}{cc}
\mathbb{1}_3 & \mathbb{0}_3\\\mathbb{0}_3 & \mathbb{1}_3
\end{array}\right)
$,
where $\mathbb{1}_3$ and $\mathbb{0}_3$ are the $3\times3$ identity matrix and zero matrix.

In the main text, we defined Bogoliubov operators such that the steady state is the vacuum of $\hb_1$ and $\hb_{i>2}$, and the dynamics preserves the quadratures of $\hb_2$. Therefore, in order to calculate $\Sigma$, it is convenient to first calculate the initial and steady state covariance matrix for the quadratures of the $\hb$ operators.

Using the definitions of $\hb_i$, i.e.
\begin{align}\label{eqn: Bogoliubov transformation X}
&\left(\begin{array}{c} \hat X^b_1 \\ \hat X^b_2 \\ \hat X^b_3 \end{array}\right) =
\left(\begin{array}{ccc} \frac{x}{\sqrt{xy\cos\phi}} & 0 & 0\\
-\tan\phi & 1 & 0\\
0 & 0 & 1
 \end{array}\right) 
\left(\begin{array}{c} \hat X^c_1 \\ \hat X^c_2 \\ \hat X^c_3 \end{array}\right),\\
\label{eqn: Bogoliubov transformation Y}
&\left(\begin{array}{c} \hat Y^b_1 \\ \hat Y^b_2 \\ \hat Y^b_3 \end{array}\right) =
\left(\begin{array}{ccc} \frac{y\cos\phi}{\sqrt{xy\cos\phi}} & \frac{y\sin\phi}{\sqrt{xy\cos\phi}} & 0\\
0 & 1 & 0\\
0 & 0 & 1
 \end{array}\right) 
\left(\begin{array}{c} \hat Y^c_1 \\ \hat Y^c_2 \\ \hat Y^c_3 \end{array}\right),
\end{align}
the initial covariance matrices for $\hat X^b_i$ and $\hat Y^b_j$ are
\begin{align}
& \left( \begin{array}{ccc}
\braket{\{ \hat X^b_1, \hat X^b_1\}} & \braket{\{ \hat X^b_1, \hat X^b_2\}} & \braket{\{ \hat X^b_1, \hat X^b_3\}} \\
\braket{\{ \hat X^b_2, \hat X^b_1\}} & \braket{\{ \hat X^b_2, \hat X^b_2\}} & \braket{\{ \hat X^b_2, \hat X^b_3\}} \\
\braket{\{ \hat X^b_3, \hat X^b_1\}} & \braket{\{ \hat X^b_3, \hat X^b_2\}} & \braket{\{ \hat X^b_3, \hat X^b_3\}}
\end{array}\right) \nonumber\\ & =
 \left( \begin{array}{ccc}
\frac{x}{y\cos\phi} & \frac{-x\tan\phi}{\sqrt{xy\cos\phi}} & 0\\
\frac{-x\tan\phi}{\sqrt{xy\cos\phi}} & \sec^2\phi & 0\\
0 & 0 & 1
\end{array}\right),
\end{align}
\begin{align}
& \left( \begin{array}{ccc}
\braket{\{ \hat Y^b_1, \hat Y^b_1\}} & \braket{\{ \hat Y^b_1, \hat Y^b_2\}} & \braket{\{ \hat Y^b_1, \hat Y^b_3\}} \\
\braket{\{ \hat Y^b_2, \hat Y^b_1\}} & \braket{\{ \hat Y^b_2, \hat Y^b_2\}} & \braket{\{ \hat Y^b_2, \hat Y^b_3\}} \\
\braket{\{ \hat Y^b_3, \hat Y^b_1\}} & \braket{\{ \hat Y^b_3, \hat Y^b_2\}} & \braket{\{ \hat Y^b_3, \hat Y^b_3\}}
\end{array}\right) \nonumber\\ & =
 \left( \begin{array}{ccc}
\frac{y}{x\cos\phi} & \frac{y\sin\phi}{\sqrt{xy\cos\phi}} & 0\\
\frac{y\sin\phi}{\sqrt{xy\cos\phi}} & 1 & 0\\
0 & 0 & 1
\end{array}\right),\\
& \braket{ \{ \hat X^b_i, \hat Y^b_j \} } = 0.
\end{align}

\Rev{
\subsection{Dynamics of the noise}
The master equations, $\partial_t \braket{\hat O} = N\Gamma\langle \hat{\mathscr{D}}^+\hat O \hat{\mathscr{D}}^- - \frac{1}{2}\hat{\mathscr{D}}^+\hat{\mathscr{D}}^-\hat O - \frac{1}{2}\hat O\hat{\mathscr{D}}^+\hat{\mathscr{D}}^-
\rangle$, for the observables $\hat O = \hat X^b_i \hat X^b_j, \hat Y^b_i \hat Y^b_j$, and $\hat X^b_i \hat Y^b_j$ are
\begin{align}
& \partial_t \braket{(\hat X^b_1)^2} = N\Gamma xy\cos\phi (1-2\braket{(\hat X^b_1)^2}), \nonumber\\
& \partial_t \braket{\hat X^b_1 \hat X^b_j} = -N\Gamma xy\cos\phi \braket{\hat X^b_1 \hat X^b_j}, \nonumber\\
& \partial_t \braket{\hat X^b_i \hat X^b_j} = 0,\nonumber\\
& \partial_t \braket{(\hat Y^b_1)^2} = N\Gamma xy\cos\phi (1-2\braket{(\hat Y^b_1)^2}), \nonumber\\
& \partial_t \braket{\hat Y^b_1 \hat Y^b_j} = -N\Gamma xy\cos\phi \braket{\hat Y^b_1 \hat Y^b_j}, \nonumber\\
& \partial_t \braket{\hat Y^b_i \hat Y^b_j} = 0,\nonumber\\
& \partial_t \braket{\hat X^b_i \hat Y^b_j} = 0,
\end{align}
where $i,j>1$. Solving them gives the time-dependent solution for the covariance matrix elements,
%\begin{widetext}
\begin{align}
\label{eqn: Bogoliubov steady state X}
& \left( \begin{array}{ccc}
\braket{\{ \hat X^b_1, \hat X^b_1\}} & \braket{\{ \hat X^b_1, \hat X^b_2\}} & \braket{\{ \hat X^b_1, \hat X^b_3\}} \\
\braket{\{ \hat X^b_2, \hat X^b_1\}} & \braket{\{ \hat X^b_2, \hat X^b_2\}} & \braket{\{ \hat X^b_2, \hat X^b_3\}} \\
\braket{\{ \hat X^b_3, \hat X^b_1\}} & \braket{\{ \hat X^b_3, \hat X^b_2\}} & \braket{\{ \hat X^b_3, \hat X^b_3\}}
\end{array}\right) =
\nonumber\\&
 \left( \begin{array}{ccc}
\frac{x}{y\cos\phi}f + (1-f) & \frac{-x \tan\phi}{\sqrt{xy\cos\phi}}f & 0\\
\frac{-x\tan\phi}{\sqrt{xy\cos\phi}}f & \sec^2\phi & 0\\
0 & 0 & 1
\end{array}\right), \\
\label{eqn: Bogoliubov steady state Y}
& \left( \begin{array}{ccc}
\braket{\{ \hat Y^b_1, \hat Y^b_1\}} & \braket{\{ \hat Y^b_1, \hat Y^b_2\}} & \braket{\{ \hat Y^b_1, \hat Y^b_3\}} \\
\braket{\{ \hat Y^b_2, \hat Y^b_1\}} & \braket{\{ \hat Y^b_2, \hat Y^b_2\}} & \braket{\{ \hat Y^b_2, \hat Y^b_3\}} \\
\braket{\{ \hat Y^b_3, \hat Y^b_1\}} & \braket{\{ \hat Y^b_3, \hat Y^b_2\}} & \braket{\{ \hat Y^b_3, \hat Y^b_3\}}
\end{array}\right) =
\nonumber\\&
 \left( \begin{array}{ccc}
\frac{y}{x\cos\phi}f + (1-f) & \frac{y\sin\phi}{\sqrt{xy\cos\phi}}f & 0\\
\frac{y\sin\phi}{\sqrt{xy\cos\phi}}e^{-N\Gamma txy\cos\phi} & 1 & 0\\
0 & 0 & 1
\end{array}\right),\\
& \braket{ \{ \hat X^b_i, \hat Y^b_j \} } = 0, \label{eqn: 0offdiagonal}
\end{align}
%\end{widetext}
where $f = e^{-N\Gamma t xy\cos\phi}$.

Inverting the Bogoliubov transformation [Eq.~\eqref{eqn: Bogoliubov transformation X} and~\eqref{eqn: Bogoliubov transformation Y}] gives $\Sigma$ in the steady state. $\Sigma$ has a block-diagonal form due to Eq.~\eqref{eqn: 0offdiagonal}, 
$\Sigma = \left( \begin{array}{cc}
\Sigma_{XX} & \mathbb{0}_3\\\mathbb{0}_3 & \Sigma_{YY}
\end{array}\right).$ 
The eigenvalues of $\Sigma_{XX}$ and $\Sigma_{YY}$ give the widths of the noise distribution.
Two of these eigenvalues are always equal to $1$ because $\hc_3$ undergoes no dynamics [see bottom right entry in Eqs.~\eqref{eqn: Bogoliubov transformation X},\eqref{eqn: Bogoliubov transformation Y},\eqref{eqn: Bogoliubov steady state X} and~\eqref{eqn: Bogoliubov steady state Y}]. The nontrivial eigenvalues in the steady state are
\begin{align}\label{eqn: squeezing}
\xi_1^2 = \frac{A_x+1}{2} - \sqrt{\left(\frac{A_x-1}{2}\right)^2 + (1-f)^2\tan^2\phi} \nonumber\\
\xi_2^2 = \frac{A_y+1}{2} - \sqrt{\left(\frac{A_y-1}{2}\right)^2 + (1-f)^2\tan^2\phi} \nonumber\\
\xi_3^2 = \frac{A_x+1}{2} + \sqrt{\left(\frac{A_x-1}{2}\right)^2 + (1-f)^2\tan^2\phi} \nonumber\\
\xi_4^2 = \frac{A_y+1}{2} + \sqrt{\left(\frac{A_y-1}{2}\right)^2 + (1-f)^2\tan^2\phi},
\end{align}
where $A_x = \frac{y}{x\cos\phi}(1-f^2) - (1-2f) + (1-f)^2\sec^2\phi$ and $A_y = \frac{x}{y\cos\phi}(1-f^2) - (1-2f) + (1-f)^2\sec^2\phi$.

The top two lines in Eq.~\eqref{eqn: squeezing} give the steady-state squeezing in the $X$ and $Y$ quadratures respectively, and the bottom two lines give the steady-state anti-squeezing. 
Note that the spin squeezing cannot be extracted by diagonalizing the covariance matrix for $\hat X^b_\mu$ [Eq.~\eqref{eqn: Bogoliubov steady state X}] and $\hat Y^b_\mu$ [Eq.~\eqref{eqn: Bogoliubov steady state Y}], since the Bogoliubov transformation is not unitary.

\subsection{Steady-state noise}
}

The steady state values of the squeezing and anti-squeezing, obtained by setting $f=0$, are
\begin{align}\label{eqn: steady squeezing}
\xi^2_1 & = \frac{ y^2\left(1+\frac{x}{y\cos\phi}\right) - y\sqrt{y^2\left(1+\frac{x}{y\cos\phi}\right)^2 - 4xy\cos\phi} } {2xy\cos\phi } , \nonumber\\
\xi^2_2 & = \frac{ x^2\left(1+\frac{y}{x\cos\phi}\right) - x\sqrt{x^2\left(1+\frac{y}{x\cos\phi}\right)^2 - 4xy\cos\phi} } {2xy\cos\phi } , \nonumber\\
\xi^2_3 & = \frac{ y^2\left(1+\frac{x}{y\cos\phi}\right) + y\sqrt{y^2\left(1+\frac{x}{y\cos\phi}\right)^2 - 4xy\cos\phi} } {2xy\cos\phi } , \nonumber\\
\xi^2_4 & = \frac{ x^2\left(1+\frac{y}{x\cos\phi}\right) + x\sqrt{x^2\left(1+\frac{y}{x\cos\phi}\right)^2 - 4xy\cos\phi} } {2xy\cos\phi }.
\end{align}

\section{Numerical Benchmarking}
We benchmark our results with cumulant and exact diagonalization simulations, as described in this section. Since both approximations go beyond the HP approximation, we also describe a procedure to obtain the squeezing of the system in the spin basis without having to explicitly define the HP bosonic variables.

\subsection{Gell-Mann matrices and spin squeezing}

For an $\ell$-level system, a complete basis of single-particle operators is given by the generalized Gell-Mann matrices $\hat g_{i,k}, k \in [1,\ell^2-1], i \in [1,N]$, defined such that ${\rm Tr}(g_{i,k}g_{i,l}) = 2\delta_{kl}$. 
For fully symmetric states, the relevant observables are \emph{collective} Gell-Mann matrices, $\hat G_k = \sum_{i=1}^N \hat g_{i,k}$.

To compute squeezing, we define the spin covariance matrix, $\tilde{\Sigma}$, with matrix elements $\tilde{\Sigma}_{lm} = \braket{ \{ \hat G_l, \hat G_m \} } - 2\braket{\hat G_l}\braket{\hat G_m}$.
The covariance matrix $\tilde\Sigma$ is an $(\ell^2-1)\times(\ell^2-1)$ matrix. However, for large $N$ and in the stable phase, we find using the cumulant approximation (see below) that only $2(\ell-1)$ eigenvalues are significant, and the others are insignificant. 
This is consistent with the expectation that the Bloch vector's length stays nearly constant, and therefore the only relevant covariances involve the $2(\ell-1)$ spin variables perpendicular to the Bloch vector. 

These orthogonal variables correspond to the Gell-Mann operators $\hat{\Lambda}_\mu = \sum_i \ket{0}_i\bra{\mu}_i + {\rm h.c.}$ ($\mu>0$) corresponding to coherences between the condensed direction $\ket{0}$ and the orthogonal states $\ket{\mu>0}$, and their variances in the large-$N$ approximation are equal to the variances of the $\hc_{\mu>0}$ bosons.
Thus, the $2(\ell-1)$ largest eigenvalues of $\tilde\Sigma$ are approximately the same as the eigenvalues of the $\Sigma$ HP covariance matrix.
Similarly, the orthogonal spin variables $\hat{S}_{\perp,\gamma}$ mentioned in the main text would correspond to linear combinations of $\hat{\Lambda}_\mu$.
For the examples considered in this paper, $\ell$ is effectively 2 (if $\beta$ is a multiple of $\pi$) or 4 (for non-multiples of $\pi$).

\subsection{Cumulant expansion}
In Fig. 3 of the main text, we plotted the noise in spin variables that were calculated using a numerical cumulant expansion. Here, we explain this method.

We write the master equation for one-point operators $\braket{\hat G_l}$ and two-point correlators $\braket{\hat G_l \hat G_m}$. The master equation for $n$-point operators involves $(n+1)$-point correlators.
Our approximation consists in setting the third-order cumulant to zero, i.e.
\begin{align}
\braket{\hat G_l \hat G_m \hat G_n} \approx &\braket{\hat G_l \hat G_m}\braket{\hat G_n} + \braket{\hat G_l \hat G_n}\braket{\hat G_m} + \braket{\hat G_m \hat G_n}\braket{\hat G_l} \nonumber\\ &- 2\braket{\hat G_l}\braket{\hat G_m}\braket{\hat G_n}.
\end{align}
This approximation works well for near-Gaussian states. Our initial uncorrelated state is Gaussian, and the many-body state stays near-Gaussian in the stable superradiant phase. Therefore, we expect our approximation to be valid in this regime. We chose $N=10^6$ for Fig.~3 in the main text.

\subsection{Exact diagonalization}
\begin{figure}[t]
\includegraphics[width=1.0\columnwidth]{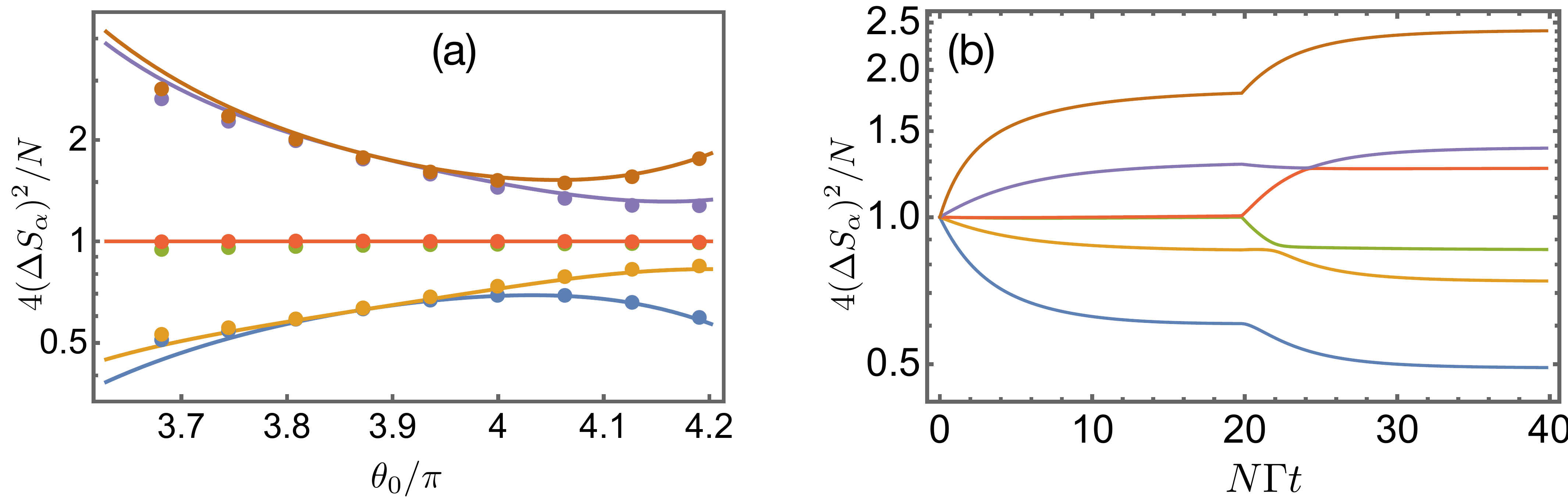}
\caption{Numerical benchmarking with exact diagonalization. (a) We plot the steady-state in the six noise quadratures perpendicular to the Bloch vector, obtained from exact diagonalization for $N=30$ as dots, and from HP as solid lines. The lines agree excellently with the dots. (b) Numerically exact simulation of the protocol to create squeezing at $(\theta_0,\beta) = (4.07\pi,0.5\pi)$ until $N\Gamma t=20$, and store it in $(\theta_{\rm dark}, \beta) \approx (3.87\pi, 0.5\pi)$.}
\label{suppFig}
\end{figure}

We further benchmark the HP predictions with numerical exact diagonalization on small systems ($N=30$). Restricting to the space of fully symmetric states, we exactly simulate the system's evolution and calculate the spins' covariance matrix $\tilde\Sigma$. Figure~\ref{suppFig}(a), which plots the $2(\ell-1)=6$ leading steady-state eigenvalues of $\tilde\Sigma$ versus $\theta_0$ at $\beta = \pi/2$, shows that the HP predictions (lines) agree excellently with the numerically exact results (dots).

Furthermore, we also numerically benchmark our protocol to create and store the squeezing in a dark state, in Fig.~\ref{suppFig}(b). We perform the spin rotation to the dark manifold at $N\Gamma t = 20$ and turn off the drive. Although the squeezing is not preserved for $N\Gamma t > 20$, due to finite-size effects ($N=30$), Fig.~\ref{suppFig}(b) shows that the system in fact gets \emph{more squeezed} after turning off the drive.

\section{Entanglement criterion}

Ref.~\cite{vitagliano2011spin} showed several tight inequalities satisfied by all separable states, violating which is a proof of entanglement. In this section, we consider one of these inequalities, and rewrite it in a form reminiscent of the spin-squeezing criterion for spin-1/2 systems. In the HP approximation, violating this inequality is equivalent to obtaining an eigenvalue of the covariance matrix $\Sigma$ to be $<1$.

For $\ell$-level systems and collective Gell-Mann matrices $\{ \hat G_k \}$, Ref.~\cite{vitagliano2011spin} proved that
\begin{equation}\label{inequality Toth}
(N-1)\sum_{k \in I} (\tilde{\Delta} G_k)^2 - \sum_{k \notin I} \braket{\tilde{G}_k^2} \geq -2N(N-1)\frac{\ell-1}{\ell}.
\end{equation}
Here, $\braket{ \tilde{G}_k^2 } = \braket{\hat G_k^2} - \sum_{i=1}^N \braket{ \hat g_{i,k}^2}$ is the second moment and $(\tilde{\Delta}G_k)^2 = (\Delta G_k)^2 - \sum_{i=1}^N \braket{\hat g_{i,k}^2}$ is the modified variance of $\hat G_k$ after subtracting same-spin terms. The set $I$ can be any subset of $[1,2,\cdots,\ell^2-1]$. Thus, there are $2^{\ell^2-1}$ inequalities, and violating any one of them is proof of entanglement. Inserting the definitions of $(\tilde{\Delta}G_k)^2 $ and $\braket{ \tilde{G}_k^2 }$, we can rewrite inequality~\eqref{inequality Toth} as
\begin{align}\label{inequality 2}
&\sum_{k \in I} \left( N (\Delta G_k)^2 - N\left( \sum_{i=1}^N \braket{\hat g_{i,k}^2}\right) + \braket{\hat G_k}^2 \right)\nonumber\\
& \geq \left(\sum_k \braket{\hat G_k^2} - \sum_{i=1}^N \braket{\hat g_{i,k}^2} \right) -2N(N-1)\frac{\ell-1}{\ell} \nonumber\\
&= \left(\sum_k \sum_{i\neq j} \braket{\hat g_{i,k}\hat g_{j,k}}\right) - 2N(N-1)\frac{\ell-1}{\ell}.
\end{align}
The term inside the parantheses in the last line is a Cassimir operator, and all collective states are eigenstates of this operator with the same eigenvalue. For two particles $i$ and $j$, the eigenvalue is $2(\ell-1)/\ell$, and summing over $i\neq j$ multiplies it by $N(N-1)$. 
Therefore, the entire right hand side of the inequality simplifies to 0. Inequality~\eqref{inequality 2} then simplifies to
\begin{equation}\label{inequality 3}
\sum_{k \in I} (\Delta G_k)^2 \geq \sum_{k\in I} \left( \sum_{i=1}^N \braket{\hat g_{i,k}^2}\right) - \frac{\braket{\hat G_k}^2}{N}.
\end{equation}

We now make a special choice of $I$ as a set with only one element. We choose this element as the Gell-Mann matrix $\hat \Lambda_\mu = \sum_i \ket{0}_i\bra{\mu}_i + {\rm h.c.} = \hc_0\+ \hc_\mu + {\rm h.c.}$, where $\ket{0}$ is the macroscopically occupied state, and $\ket{\mu}$ is orthonormal to $\ket{0}$. The term $\sum_{i=1}^N \braket{\hat g_{i,k}^2}$ simplifies to $n_0 + n_\mu$, where $n_0$ and $n_\mu$ are populations in $\ket{0}$ and $\ket{\mu}$. In this case, inequality~\eqref{inequality 3} is 
\begin{equation}
(\Delta \Lambda_\mu)^2 \geq n_0 + n_\mu - \frac{\braket{\hat \Lambda_\mu}^2}{N} \geq n_0 - \frac{\braket{\hat \Lambda_\mu}^2}{N}
\end{equation}
where the second inequality is tighter, and follows from $n_\mu \geq 0$. Violating these inequalities is proof of entanglement.

In the HP approximation, $\braket{\hat \Lambda_\mu} \approx \sqrt{2N}\braket{\hat X^c_\mu} \approx 0$, $n_0 = \braket{\hc_0\+ \hc_0^{\phantom\dagger}} \approx N$, and $(\Delta \Lambda_\mu)^2 = 2N(\Delta X^c_\mu)^2$. If we choose $\ket{\mu}$ such that $X^c_\mu$ is an eigenvector of $\tilde\Sigma$, then its eigenvalue is $\xi^2_\mu = 2(\Delta X^c_\mu)^2$. Thus, the squeezing inequality reduces to $\xi^2_\mu\geq1$, i.e.~having an eigenvalue $<1$ indicates entanglement. Replacing $\ket{\mu}$ with $i\ket{\mu}$ gives $(\Delta \Lambda_\mu)^2 = 2N(\Delta Y^c_\mu)^2$.

\section{Dark state close to the critical point}
In the main text, we considered a case where squeezing is directly created in a dark state via collective decay. For this we initialized the atoms in $\ket{\Psi(\theta_0;\beta)} = \exp( -i\theta_0\hat D^x_{+1}) \allowbreak \left( \cos\beta\ket{g,-\frac{1}{2}} + \sin\beta\ket{g,\frac{1}{2}}\right)^{\otimes N}$, where we chose $\beta$ such that this state is a mean-field dark state and it is close to a critical point. Here, we derive the condition for such $\beta$ and $\theta_0$.

The dark state and critical point conditions are, respectively,
\begin{align}
& \frac{\partial V_\beta(\theta)}{\partial\theta}\vert_{\theta=\theta_0} = \frac{\cos^2(\beta/2)}{2\sqrt{3}}\sin\frac{\theta}{\sqrt{3}} + \frac{\sin^2(\beta/2)}{2}\sin\theta = 0, \nonumber\\
& \frac{\partial^2 V_\beta(\theta)}{\partial\theta^2}\vert_{\theta=\theta_0} = \frac{\cos^2(\beta/2)}{6}\cos\frac{\theta}{\sqrt{3}} + \frac{\sin^2(\beta/2)}{2}\cos\theta = 0.
\end{align}
Both equations will be satisfied if $\tan\frac{\theta_0}{\sqrt{3}} = \frac{\tan\theta_0}{\sqrt{3}}$ and $\beta = 2\tan^{-1}\sqrt{ \frac{-\cos(\theta_0/\sqrt{3})}{3\cos\theta_0} }$. The parameters $\theta_0 \approx 2.45\pi$ and $\beta \approx 0.41\pi$ used in the main text satisfy these equations.

\section{Transferring the squeezing}

We specify here the rotations employed for the case where squeezing is created in a bright state close to a critical point and then transferred to a dark state.

As a first step, upon reaching the steady state, we rotate the Bloch vector to the dark state. For this we use the same external drive that we used for the initial state preparation and continuous driving. This operation also rotates the noise quadratures along with it. In general, when the Bloch vector rotates from any angle $\theta_1$ to $\theta_2$ via a rotation $\exp[ i(\theta_1-\theta_2)\hat D^x_{+1}]$, the map for the rotation of the noise directions is given by
\begin{align}
& \hc_1 \rightarrow \hc_1, \nonumber\\
& \hc_2 \rightarrow \lambda \hc_2 + \lambda' \hc_3, \nonumber\\
& \hc_3 \rightarrow \lambda' \hc_2 - \lambda \hc_3,
\end{align}
For $\beta = 0.5\pi$, $\lambda$ and $\lambda'$ are given by $\lambda = \vec{v}_1(\theta_1)\cdot \vec{v}_1(\theta_2), \lambda' = \vec{v}_1(\theta_1)\cdot \vec{v}_2(\theta_2)$ where 
\begin{widetext}
\begin{align}
& \vec{v}_1 = \frac{1}{2y\sin\phi}\left( \frac{\sin\theta - \frac{1}{\sqrt{3}}\sin\frac{\theta}{\sqrt{3}} }{\sqrt{2}}, \frac{1}{\sqrt{3}}\left(\cos\frac{\theta}{\sqrt{3}} - \frac{y\cos\phi}{x}\right), \cos\theta - \frac{y\cos\phi}{x}\right), \nonumber\\
& \vec{v}_2 = \frac{1}{4xy\sin\phi}\left( \frac{\cos\theta - \cos\frac{\theta}{\sqrt{3}}}{\sqrt{3}}, \frac{\sin\theta - \frac{1}{\sqrt{3}}\sin\frac{\theta}{\sqrt{3}}}{\sqrt{2}}, \frac{ \frac{1}{\sqrt{3}}\sin\frac{\theta}{\sqrt{3}} - \sin\theta}{\sqrt{6}}\right)
\end{align}
\end{widetext}
In our case, we rotate from $\theta_1 \sim \theta_c$ to $\theta_2 = \theta_{\rm dark}$. Note that in principle we can rotate to any of the dark states $\theta_{\rm dark}$ available in the superradiance potential.

\begin{figure}[t]
\includegraphics[width=1.0\columnwidth]{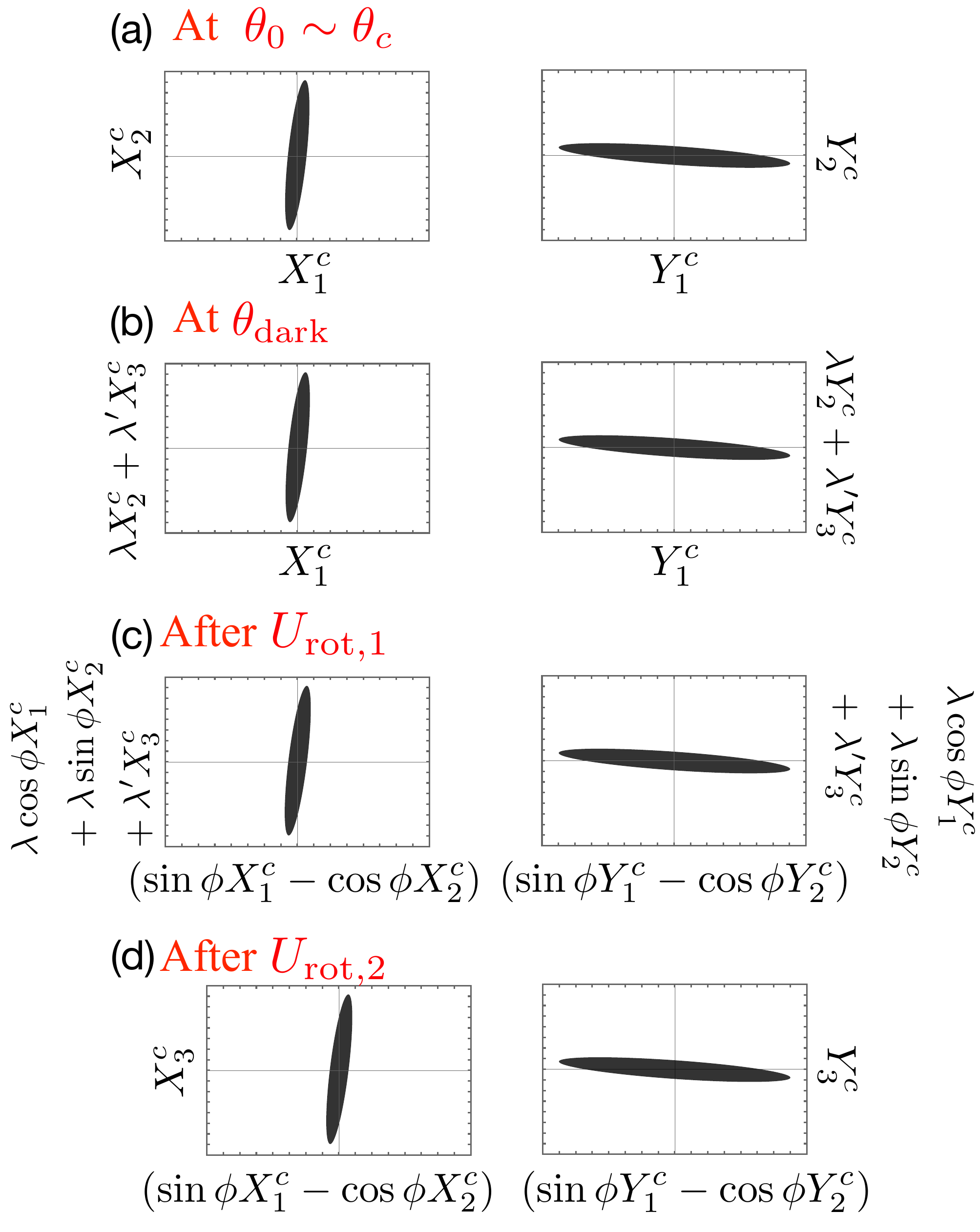}
\caption{Illustration of noise distributions during the protocol to transfer the squeezing to the dark state. (a) The system is squeezed along $X^c_1$ and $Y^c_2$ at $\theta_0 \sim \theta_c$. (b) The squeezed quadratures are $X^c_1$ and $\lambda Y^c_2 + \lambda' Y^c_3$ immediately after rotating to the dark state. (c) After applying $U_{\rm rot,1}$, the squeezed quadratures are $\sin\phi X^c_1 - \cos\phi X^c_2$ and $\lambda(\cos\phi Y^c_1 + \sin\phi Y^c_2) + \lambda' Y^c_3$. After this, the squeezing in $\sin\phi X^c_1 - \cos\phi X^c_2$ will be preserved. (d) After applying $U_{\rm rot,2}$, the squeezed Y quadrature is $Y^c_3$, and will be preserved after this.}
\label{suppFig2}
\end{figure}

When the system settles into the steady state close to the critical point, the squeezed quadratures are close to $X^c_1$ and $Y^c_2$, and the anti-squeezed quadratures are close to $X^c_2$ and $Y^c_1$ [see Fig.~\ref{suppFig2}(a)].
Thus, from the above expressions we see that immediately after rotating to the dark state, the squeezed quadratures are $X^c_1$ and $\lambda Y^c_2 + \lambda' Y^c_3$, and the anti-squeezed quadratures are $\lambda X^c_2 + \lambda' X^c_3$ and $Y^c_1$ [see Fig.~\ref{suppFig2}(b)].

As a second step, we would like to rotate the squeezed quadratures to align along the conserved directions.
First, we describe how to rotate $X^c_1$ to $\sin\phi_{\rm dark} X^c_1 - \cos\phi_{\rm dark} X^c_2 \propto X^b_2(\theta_{\rm dark})$, where $\theta_{\rm dark}$ is associated with a dark state and $\phi_{\rm dark} = \phi(\theta_{\rm dark})$. The rotation of the bosonic operator $\hc_i$ to $\hc_j$, where $\hc_i$ and $\hc_j$ are not necessarily independent, is achieved by the beam-splitter $\hat H_{\rm rot} = i(\hc_i\+ \hc_j - \hc_j\+ \hc_i)$. Time evolution of the operators under this Hamiltonian is given by
\begin{align}
& \hc_i(\tau) = \hc_i \cos(\tau\sqrt{1-\lambda^2}) + \frac{(\hc_j-\lambda\hc_i)}{\sqrt{1-\lambda^2}} \sin(\tau\sqrt{1-\lambda^2}), \nonumber\\
& \hc_j(\tau) = \hc_j \cos(\tau\sqrt{1-\lambda^2}) - \frac{(\hc_i-\lambda\hc_j)}{\sqrt{1-\lambda^2}} \sin(\tau\sqrt{1-\lambda^2}),
\end{align}
where $\lambda = [\hc_i, \hc_j\+]$. Thus, $\hc_i(\tau) = \hc_j$ at $\tau = \frac{\cos^{-1}\lambda}{\sqrt{1-\lambda^2}}$. The desired rotation from $X^c_1$ to $\sin\phi_{\rm dark} X^c_2 - \cos\phi_{\rm dark} X^c_1$ is accomplished by $U_{\rm rot,1} = \exp\left(\left(\frac{\pi}{2}-\phi_{\rm dark}\right) (\hc_1\+ \hc_2 - \hc_2\+ \hc_1)\right)$.

Immediately after $U_{\rm rot,1}$, the squeezed quadratures are $\sin\phi_{\rm dark} X^c_1 - \cos\phi_{\rm dark} X^c_2$ and $\lambda(\cos\phi_{\rm dark} Y^c_1 + \sin\phi_{\rm dark} Y^c_2) + \lambda' Y^c_3$, and the anti-squeezed quadratures are $\lambda(\cos\phi_{\rm dark} X^c_1 + \sin\phi_{\rm dark} X^c_2) + \lambda' X^c_3$ and $\sin\phi_{\rm dark} Y^c_1 - \cos\phi_{\rm dark} Y^c_2$ [see Fig.~\ref{suppFig2}(c)].

Next, we wish to rotate $\lambda(\cos\phi_{\rm dark} Y^c_1 + \sin\phi_{\rm dark} Y^c_2) + \lambda' Y^c_3$ to $Y^c_3$. This can also be achieved using an operation $U_{\rm rot,2}$ derived from the recipe given above. After this rotation, the squeezed quadratures are $\sin\phi_{\rm dark} X^c_1 - \cos\phi_{\rm dark} X^c_2$ and $Y^c_3$, and the anti-squeezed quadratures are $X^c_3$ and $\sin\phi_{\rm dark} Y^c_1 - \cos\phi_{\rm dark} Y^c_2$ [see Fig.~\ref{suppFig2}(d)].

For $\beta = 0.5\pi$, $\theta_c = 4.47\pi$, and $\theta_{\rm dark} = 3.87\pi$, 
the two rotations have the form $U_{\rm rot,1} = \exp(i \hat H_{eg}^{(1)})$ and $U_{\rm rot,2} = \exp(i (\hat H_g^{(2)} + \hat H_e^{(2)} + \hat H_{eg}^{(2)}))$, where
\begin{align}\label{eqn: rotation Hamiltonians}
&\hat H_{eg}^{(1)} = -0.74\hat S^x_{-1/2} + 1.28 \hat S^x_{1/2} - 0.71\hat \Pi^x_{1/2} + 1.36\hat T^x_{-1/2}\nonumber\\
&\hat H_g^{(2)} = -0.12\hat S^y_g, \nonumber\\
&\hat H_e^{(2)} = -1.58\hat S^y_e, \nonumber\\
&\hat H_{eg}^{(2)} = 
-0.42\hat S^x_{-1/2} - 0.44\hat S^x_{1/2} - 0.08\hat \Pi^x_{1/2} + 0.14\hat T^x_{-1/2}.
\end{align}
$\hat S^\alpha_{g}$ and $\hat S^\alpha_e$ are spin-1/2 operators within the ground manifold, and within the two-level excited manifold $\{\ket{e,1/2}, \ket{e,3/2}\}$, $\hat S^\alpha_m$ are the spin-1/2 operators within the manifold $\{\ket{g,m},\ket{e,m+1}\}$ as defined in the main text, and $\hat \Pi^x_{1/2} = \frac{1}{2}\ket{g,1/2}\bra{e,1/2} + {\rm h.c.}$ and $\hat T^x_{-1/2} = \frac{1}{2}\ket{g,-1/2}\bra{e,3/2} + {\rm h.c.}$. The Hamiltonians in Eq.~\eqref{eqn: rotation Hamiltonians} can potentially be implemented via a sequence of magnetic fields, Stark shifts, and microwave and Raman transitions. As a guiding example, below we also provide an explicit decomposition of $U_{\rm rot,1}$ which may be implemented in an experiment:
\begin{align}
& U_{\rm rot,1} = (R_g R_e)\+ \times R_{eg}\times (R_g R_e),\nonumber\\
& R_g = \exp\left( i\pi(0.69\hat S^x_g + 0.72\hat S^z_g)\right),\nonumber\\
& R_e = \exp\left( i\pi(0.48\hat S^z_e - 0.88\hat S^x_e)\right),\nonumber\\
& R_{eg} = \exp\left( -0.68i \pi \hat S^x_{-1/2}\right).
\end{align}
We point out that while the separate steps $R_g, R_e$ and $R_{eg}$ may rotate the Bloch vector away from the dark state, the sequence of rotations which compose $U_{\rm rot,1}$ will return the Bloch vector to the dark state.

\Rev{
\section{Effect of single-particle decoherence}
In the main text, we considered collective dissipation as the only source of decoherence, and studied how the interplay of driving and collective dissipation can lead to squeezing. However, atoms can also spontaneously emit light into free space, which is a source of decoherence which degrades entanglement and squeezing. The jump operators for this process are 
\begin{equation}
\hat d^-_{i,\alpha} = \sqrt{\gamma}\sum_m C_m^\alpha \hat s^-_{m,i,\alpha}
\end{equation}
where $\alpha$ denotes the polarization of the emission, $\gamma$ is the linewidth of the excited state, and $\hat s^-_{m,i,\alpha}$ is the single-particle spin lowering operator for the transition between $\ket{e,m+\alpha}$ and $\ket{g,m}$.

The degradation of the squeezing due to spontaneous emission is 
due to additional corrections which typically take the form $\xi_\mu^2 = \xi_{\mu,0}^2 + \tilde{\gamma} t$, with $\xi_{\mu,0}^2$ the squeezing due to only collective dissipation, and $\tilde{\gamma} \propto \gamma$. The expression for $\xi_{\mu,0}^2$ was given in Eq.~\eqref{eqn: squeezing}. It monotonically decreases from 1 as $t$ varies from 0 to $\infty$, whereas $\tilde{\gamma} t$ monotonically increases. Thus, there is an optimal time where the squeezing is optimal. We calculate this below, and consider only $\xi_1^2$ for concreteness. A similar analysis can be applied for $\xi_2^2$.

\subsection{The optimal value of squeezing}
We minimize the squeezing by solving 
\begin{align}\label{eqn: squeezing derivative}
\frac{d\xi_{1,0}^2}{dt} + \tilde{\gamma} = 0.
\end{align}
Eq.~\eqref{eqn: squeezing derivative} has a solution in the regime $\cos\phi \ll 1-f\ll 1$. In this time window,
\begin{equation}\label{eqn: squeezing derivative_t}
\frac{d\xi_{1,0}^2}{dt} \approx -2N\Gamma y^2 \frac{\cos^2\phi}{(1-f)^2} + O\left(\frac{\cos^3\phi}{(1-f)^3}\right),
\end{equation}
therefore the solution for Eq.~\eqref{eqn: squeezing derivative} is $(1-f)/\cos\phi \approx y\sqrt{2NC}$ where $C = \Gamma/\tilde{\gamma}$ is proportional to the single-particle cavity cooperativity. Furthermore, since $1-f\ll 1$, we can Taylor expand the exponent in $f$, $f \approx 1-N\Gamma t xy \cos\phi$, and thus obtain the optimal time for the squeezing,
\begin{equation}\label{eqn: optimal time}
t \approx \frac{1}{\Gamma x}\sqrt{\frac{2C}{N}}.
\end{equation}
The value of the squeezing at this optimal time is
\begin{equation}
\xi_1^2 \approx \frac{2}{x}\sqrt{\frac{2}{NC}}.
\end{equation}
The condition $(1-f)\approx N\Gamma t xy\cos\phi \ll 1$ also means that the equation for the optimal time [Eq.~\eqref{eqn: optimal time}] is valid only if $\cos\phi \ll \frac{1}{y\sqrt{2NC}}$.

\section{Higher-order corrections in $N$}
In the main text, we approximated the jump operator as $\hat{\mathscr{D}}^-_{+1} \approx \sqrt{N}(x \hat{X}_1 + iy(\cos\phi \hat{Y}_1 + \sin\phi \hat{Y}_2))$, and we ignored corrections of O(1). The O(1) corrections modify the value of the squeezing, and in particular, limit the best squeezing achievable. We observed this limiting in Fig.~3(d), where we plotted the best squeezing that was numerically computed from solving the coupled equations in the cumulant expansion. Here, we give a brief analytical argument that supports the numerical scaling observed for the best squeezing. We also compare it with the best squeezing achievable when spontaneous emission is also present.

To find the best squeezing and the time at which the best squeezing is achieved, we solve $\frac{d\xi^2}{dt} = 0$. The leading term in $\frac{d\xi^2}{dt}$ is $\frac{d\xi^2_0}{dt} = O(N\Gamma)$ [see e.g. Eq.~\eqref{eqn: squeezing derivative_t}], and arises from $\hat{\mathscr{D}}^-_{+1} \approx \sqrt{N}(x \hat{X}_1 + iy(\cos\phi \hat{Y}_1 + \sin\phi \hat{Y}_2))$. The next term in $\frac{d\xi^2}{dt}$ arises from the O(1) corrections in $\hat{\mathscr{D}}^-_{+1}$, and is $O(\Gamma)$. Near the critical point, the $O(\Gamma)$ contribution is nearly equal to $\Gamma \xi^2_{\rm anti-sq}$, where $\xi^2_{\rm anti-sq} = \xi^2_3$ or $\xi^2_4$ is the anti-squeezing [Eq.~\eqref{eqn: squeezing}]. A detailed derivation of this heuristic argument is left to future work~\cite{pra}. In sum, including this correction, we solve $\frac{d\xi^2}{dt} = \frac{d\xi^2_0}{dt} + \Gamma \xi^2_{\rm anti-sq} = 0$. This equation has a solution at the time $t$ that fulfils $\cos\phi \ll 1-f \ll 1$. Specifically,
\begin{equation}
\frac{d\xi^2}{dt} \approx -2N\Gamma y^2\frac{\cos^2\phi}{(1-f)^2} + \Gamma \frac{(1-f)^2}{\cos^2\phi}.
\end{equation}
The solution is $(1-f)/\cos\phi \approx (2Ny^2)^{1/4}$, and the squeezing at these optimal parameters scales as $\xi^2 \propto N^{-1/4}$.

In Fig.~3(d), we plotted the best steady-state ($t \rightarrow \infty$) squeezing, i.e., $f=0$. Thus, the best steady-state squeezing is obtained at $\cos\phi \propto N^{-1/4}$.

\subsection{Including both higher-order corrections and spontaneous emission}
In the regime $\cos\phi \ll 1-f \ll 1$,
\begin{equation}
\frac{d\xi^2}{dt} \approx \tilde{\gamma} -2N\Gamma y^2\frac{\cos^2\phi}{(1-f)^2} + \Gamma \frac{(1-f)^2}{\cos^2\phi}.
\end{equation}
The solution to $\frac{d\xi^2}{dt} = 0$ is dominated by spontaneous emission, and the squeezing scales $\propto 1/\sqrt{NC}$, if $NC^2 \ll 1$. The solution to $\frac{d\xi^2}{dt} = 0$ is dominated by finite-N corrections, and the squeezing scales $\propto N^{-1/4}$, if $NC^2 \gg 1$.
}

\section{Superradiant emission of two polarizations}
In the main text, we considered squeezing generated by the driven-dissipative dynamics for atoms populating $\ket{g,\pm 1/2}$, $\ket{e,1/2}$, and $\ket{e,3/2}$. In this case, only the right-handed polarization was relevant. The cavity can, however, support two polarization modes, and both polarizations will be relevant for the case where atoms populate all six levels, or an arbitrary initial state for atoms with an arbitrary internal structure. 
For concreteness, we focus on the six-level system from the main text, and explain the resultant squeezing.

In the case of two polarizations, the master equation for the atoms is $\hbar\dot{\rho} = \sum_\alpha \mathcal{L}_\alpha[\rho]$, where
\begin{align}
& \mathcal{L}_\alpha[\rho] = \hbar\Gamma \left( \hat{\mathscr{D}}^-_\alpha \rho \hat{\mathscr{D}}^+_\alpha - \frac{1}{2}\{ \hat{\mathscr{D}}^+_\alpha \hat{\mathscr{D}}^-_\alpha, \rho\}\right) \nonumber\\
& \hat{\mathscr{D}}^-_\alpha = \hat D^-_\alpha + i\frac{\Omega_\alpha}{\Gamma}.
\end{align}
We assume the driving strengths are chosen to keep the mean Bloch vector stationary, $\Omega_\alpha = \Gamma \braket{ \hat D^y_\alpha}$.

When the atoms populate all six levels $\ket{g,-1/2\leq m\leq 1/2}$ and $\ket{e,-3/2 \leq m\leq 3/2}$, $\ell=6$ and therefore there are six $\hc_\mu$ Schwinger bosons, and five Bogoliubov bosons $\hb_\mu$. We define the Bogoliubov bosons as follows.
First, we define Bogoliubov operators $\hb_1 \propto \hat{\mathscr{D}}^-_{+1}$ for the jump operator for right-handed polarization, and $\hb_3 \propto \hat{\mathscr{D}}^-_{-1}$ for the jump operator for left-handed polarization. We note that $\hat{b}_1$ and $\hat{b}_3$ defined in this way may not satisfy $[\hat b_1, \hat b_3\+] = 0$ (the actual Bogoliubov bosons will be linear combinations thereof). However, the steady state will still be the vacuum with respect to both $\hat{b}_1$ and $\hat{b}_3$, so the same procedure used for the one-polarization case can be applied here to find squeezing.

As before, we define $\hc_0$ to annihilate particles in $\ket{\psi(t=0)}$, and we define $\hc_1$ and $\hc_2$ as explicitly described earlier to yield the following form of the jump operator,
\begin{equation}\label{eqn: alpha=1}
\hb_1 \propto \hat{\mathscr{D}}^-_{+1} = x_1 \hat X^c_1 + i y_1 \left( \cos\phi \hat Y^c_1 + \sin\phi \hat Y^c_2\right),
\end{equation}
for some value of the parameters $x_1,y_1,\phi$.
The proportionality constant should be chosen to enforce the right commutation for $\hb_1$ and $\hb_1\+$. Since $\hat{\mathscr{D}}^-_{+1}$ and $\hat{\mathscr{D}}^-_{-1}$ commute at leading order in the HP approximation, i.e. $\braket{ [\hat{\mathscr{D}}^-_1, \hat{\mathscr{D}}^-_{-1}] }_{MF} = 0$, then $\hb_3$ necessarily has to be of the form
\begin{align}
\hb_3 \propto\ & z_1\hb_1 + x_2 (\cos\phi\hat X^c_2 - \sin\phi\hat X^c_1) + i y_2 \hat Y^c_2 \nonumber\\
&+ x_3 \hat X^c_3 + i(y_3 \hat Y^c_3 + y_4 \hat Y^c_4)
\end{align}
for some value of the parameters $z_1,x_2,y_2,x_3,y_3,y_4$,
and for an appropriate choice of $\hc_3$ and $\hc_4$, which can be defined in a similar procedure to how $\hc_1$ and $\hc_2$ were chosen.

The three other Bogoliubov bosons are defined such that they commute with $\hb_\mu$ and $\hb_\mu\+, \mu\in\{1,3\}$. For example,
\begin{align}
\hb_2 \propto &(y_3\sin\phi \hat X^c_1 - y_3\cos\phi \hat X^c_2 + y_2\cos\phi\hat X^c_3) \nonumber\\ & + i(x_2y_3\cos\phi \hat Y^c_3 - x_3y_3\hat Y^c_2 + x_2y_4\cos\phi\hat Y^c_4) \nonumber\\
\hb_4 \propto &(y_4\hat X^c_3 - y_3\hat X^c_4) + i\hat Y^c_4,\nonumber\\
\hb_5 = &\hc_5.
\end{align}

The above judicious definitions of $\hc_\mu$ and $\hb_\mu$ lets us extend our argument in the main text to describe the squeezing with two polarizations. Both $\hb_2$ and $\hb_4$ are conserved, since they commute with $\hat{\mathscr{D}}^\pm_{\pm 1}$. Due to similar arguments as the main text, conservation of $\hat X^b_2$ and $\hat X^b_4$ will lead to squeezing in at most two modes in the $X^c_1$-$X^c_2$-$X^c_3$-$X^c_4$ hyper-plane, and conservation of $\hat Y^b_2$ and $\hat Y^b_4$ will lead to squeezing of at most two modes in the $Y^c_1$-$Y^c_2$-$Y^c_3$-$Y^c_4$ hyper-plane. Thus, there can exist up to four squeezed quadratures due to dissipation into two cavity polarization modes.

\bibliography{bibliography}